\documentclass{nature}

\usepackage{lineno}
\linenumbers*[0] 
\nolinenumbers

\usepackage{amsfonts}
\usepackage{amssymb}
\usepackage{amsmath}
\usepackage{xcolor}

\usepackage{float}
\usepackage{graphicx}

\usepackage[export]{adjustbox}

\usepackage[nomarkers,nolists,figuresonly]{endfloat}

\usepackage{caption}
\usepackage[percent]{overpic}
\usepackage{subcaption}
\usepackage{hyperref}
\makeatletter
\let\saved@includegraphics\includegraphics
\AtBeginDocument{\let\includegraphics\saved@includegraphics}
\renewenvironment*{figure}{\@float{figure}}{\end@float}
\makeatother

\begin{document}

\title{Pausing ultrafast melting by timed multiple femtosecond-laser pulses}

\author{Tobias Zier$^{1,2,3}$, Eeuwe S.\ Zijlstra$^{2,3}$, Martin E.\ Garcia$^{2,3}$, David A. Strubbe$^{1}$}

\maketitle

\begin{affiliations}
 \item Department of Physics, University of California Merced, Merced, CA 95343
 \item Theoretical Physics, University of Kassel, Heinrich-Plett-Str.\ 40, 34132 Kassel, Germany
 \item Center for Interdisciplinary Nanostructure Science and Technology (CINSaT),
       Heinrich-Plett-Str.\ 40, 34132 Kassel, Germany
\end{affiliations}


\begin{abstract}
An intense femtosecond-laser excitation of a solid induces highly nonthermal conditions. 
In materials like silicon, laser-induced bond-softening leads 
to a highly incoherent ionic motion and eventually nonthermal melting. But is this 
outcome an inevitable consequence, or can it be controlled? Here, we performed 
\textit{ab initio} molecular dynamics simulations of crystalline silicon after 
timed multiple femtosecond-laser pulse excitations with fluence above the nonthermal 
melting threshold. Our results demonstrate an excitation mechanism that pauses 
nonthermal melting and creates a metastable state instead, with an electronic structure 
similar to the ground state. This mechanism can be generalized to other materials, 
potentially enabling structural and/or electronic transitions to metastable phases 
in the high-excitation regime. In addition, our approach could be used to switch off 
nonthermal contributions in experiments, allowing reliable electron-phonon coupling 
constants to be obtained more easily.

\end{abstract}


\section*{Introduction}
Intense femtosecond-laser excitations of solids lead to a variety of phenomena in 
both the electronic system and the crystalline structure. The extreme non-equilibrium 
conditions within the electronic system, or between the electronic system and the 
ions, can be used to manipulate solid-state properties and/or induce new ones far 
from equilibrium. The direct interaction of intense light pulses with the electronic 
system can be used, e.g., to tune the nonlinearity of optical properties 
by engineering Floquet states \cite{Shan_21}, to generate high harmonics 
\cite{Lv_21,Apostolova21,Luu18,Ghimire19,Yang19} or to induce non-equilibrium 
quantum phase transitions \cite{Li_21, Bandyopadhyay_21, DeNicola_21}. 
In general, laser-induced changes in the electronic system have a direct 
influence on the bonding properties and enable the ions to follow  
pathways that are forbidden or at least hard to reach in thermal equilibrium.
Recent time-resolved diffraction experiments using X-rays 
\cite{Li17,Bengtsson20} and/or electron pulses \cite{Harb08}, in combination with 
theoretical works \cite{Stampfli90,Gambiraso00,Recoules06,Zier2017}, have allowed 
identification and understanding of ensuing ultrafast structural phenomena such as 
coherent phonons \cite{Silvestri85,Zeiger92,Riffe07,Liu_21}, thermal phonon squeezing 
\cite{Johnson09,PRX2013}, ultrafast solid-to-solid phase transitions 
\cite{Her98,Cavalleri01,Rapp15,Potemkin2022} and nonthermal or ultrafast melting 
\cite{Shank83,Tom88,Saeta91,Sokolowski95,Silvestrelli96,Rousse01,Sciaini09,Hada15,
Lian16}. 

\begin{figure}[ht!]
  \centering
 \includegraphics[width=0.63\textwidth]{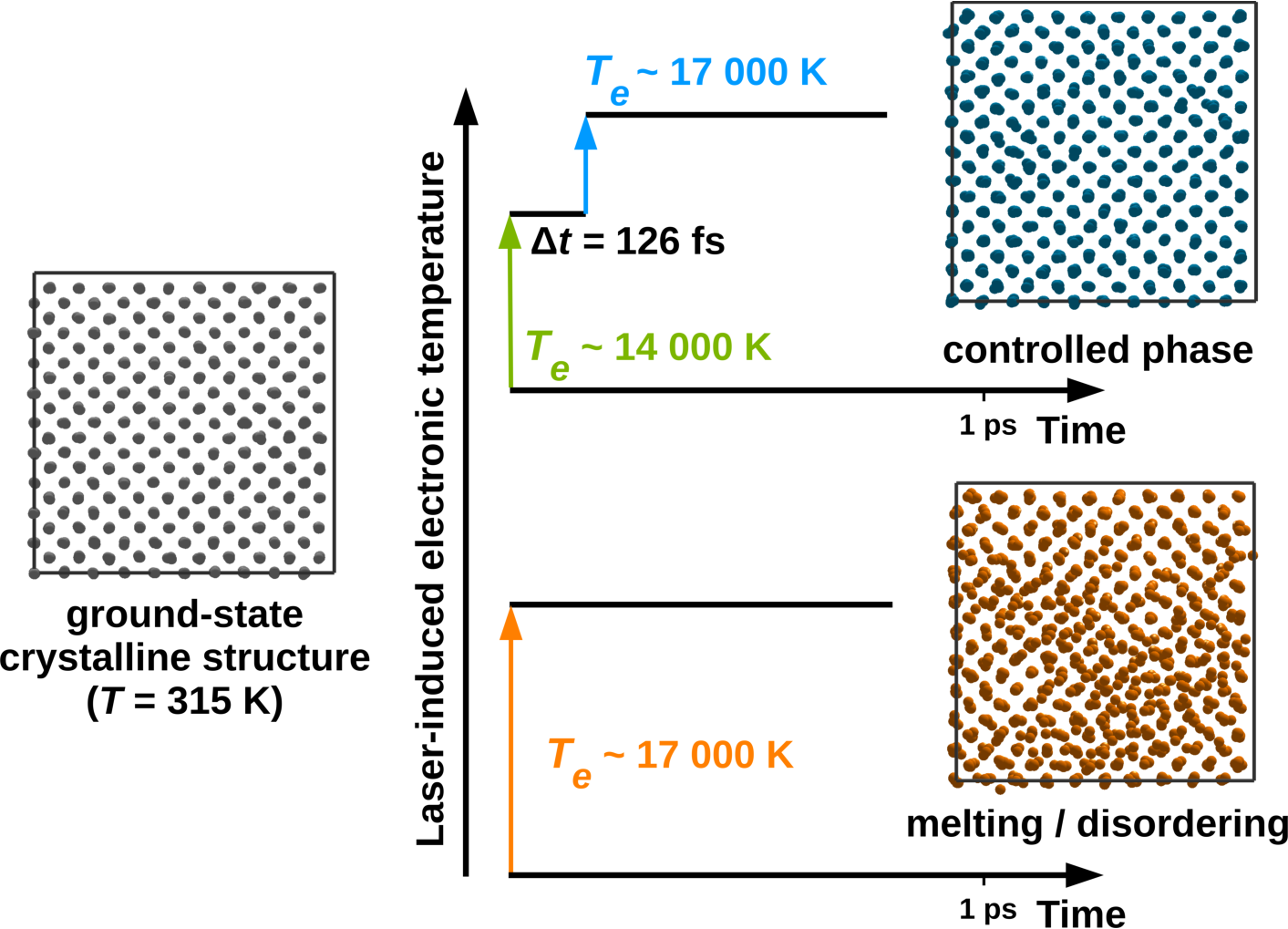}
 \captionof{figure}{\textbf{Excitation scheme:} Excitation scheme with simulation snapshots 
 for the equilibrium  structure (left) and $1$ ps after the laser excitation (right). A double pulse (right, top)  to an initial electronic temperature of $T_{\rm e} = 17 251$ K leads to a metastable state, whereas  a single pulse to $T_{\rm e} = 17 368$ K (right, bottom) with the same electronic entropy 
 $S_{\rm e} = 6.14$ mHa K$^{-1}$ causes disorder. The time delay is $\Delta t = 126$ fs.}
\label{fig:scheme}
\end{figure}

However, it remains still a challenge to distinguish thermal from nonthermal 
contributions to the overall laser-induced phenomenon. It took almost 40 years 
to solve the puzzle of whether the ultrafast laser-induced disordering process 
in silicon is of thermal or nonthermal origin \cite{VANVECHTEN1979, Silvestrelli96, 
Rousse01, Zier15, Lian16}. In the case of pure laser-induced thermal melting 
\cite{Sokolowski95, Shugaev2020}, energy is incoherently transferred from the highly 
excited electron system to the ions by electron-phonon scattering events, which 
will heat the ionic system \cite{Shugaev2020,Waldecker17}. This increase in thermal 
energy of the ions leads to ions overcoming the interatomic bonding and results in 
a disordering of the crystalline structure \cite{Sokolowski95, Shugaev2020}. 
This thermal disordering of the structure should be stochastic and independent of the 
heating process itself, as long as an equivalent thermal energy in the final state 
is reached. By contrast, in nonthermal melting the change in interatomic bonding 
due to the non-equilibrated electrons is so extreme that irreversible structural 
changes are induced \cite{Sokolowski95,Silvestrelli96,Zier15, Zier_PRL} even without 
incoherent energy transfer from the electrons to the ionic system. In such cases, 
ionic coherences can be preserved or even induced, which could be used to control 
and/or modify material properties. After a femtosecond-laser excitation both thermal 
and nonthermal effects are present, e.g., laser-induced melting in aluminum was 
found to be thermal \cite{Siwick03} in general but could show certain nonthermal 
signatures \cite{Li17thermal}. However, the timescales \cite{Zier15,CHEN05,Harb06} 
on which both effects act has been controversial. It is still a challenge in current 
research to obtain accurate electron-phonon descriptions from first principles 
\cite{Sadavisam,Ogitsu2018} and experiments \cite{Zhou2021,Jo2022, Gilberto} in order to 
sufficiently describe incoherent electron-phonon energy transfer  -- in particular, 
when both thermal and nonthermal contributions are present \cite{Jo2022}. 
As a result, a scheme to pause nonthermal contributions, enabling measurement of thermal 
contributions alone, can help unravel this complexity. 
In this paper, we investigated an excitation mechanism (Fig.\ \ref{fig:scheme}) that 
allows the pausing of nonthermal melting, using \textit{ab initio} molecular-dynamics (MD) 
simulations. The excitation scheme itself is based on a general physical 
idea, namely, the oscillating behavior of atomic kinetic and potential energy in all materials 
that show phonon squeezing \cite{Johnson09,PRX2013,Trigo2013} after an ultrashort-laser 
excitation. Our results indicate that a metastable crystalline state can be reached 
by a timed multiple-pulse excitation to a laser-induced electronic temperature at which 
normally nonthermal melting would destroy the crystalline order within a few hundreds 
of femtoseconds after excitation. The electronic structure of this metastable state 
remains similar to the ground state but with a smaller band gap. Both order and gap 
would disappear at this electronic temperature if induced using only one pulse. Our 
results show the possibility using higher-intensity laser pulses to induce structural and/or 
electronic transitions, like in nonlinear phononics \cite{Foerst2011}, without causing 
disordering. Moreover, pausing nonthermal effects offers a route to obtaining reliable 
incoherent electron-phonon coupling from experiments \cite{Waldecker17,Jo2022,Hu2022}, 
which has been a major challenge also in theory \cite{Simoni19, Liu22, Tong21}.

\section*{Results and Discussions}

\subsection{\textit{Ab initio} MD simulations}

We established the presented data by performing \textit{ab initio} MD simulations 
using CHIVES \cite{Zijlstra_09,Grigoryan2014,Zier2017}. We use an approach in which 
the femtosecond-laser excitation is modelled by an instantaneous increase of the electronic 
entropy $S_{\rm e}$, which corresponds to an energy absorption on a timescale less than our ionic 
timestep of $2$ fs, with no direct laser effect on the ions. 
On the fly, the corresponding electronic temperature $T_{\rm e}$ is computed, as well as its Fermi 
distribution, which is used to determine the electronic occupations. 
Here, we assume that a single equilibrated Fermi distribution is sufficient 
to describe the whole system, rather than a quasi-Fermi level splitting \cite{Comp_elect}. 
In silicon, experiments suggest that a common carrier temperature is rapidly established 
($\sim$ 4 fs) after an ultrashort-laser pulse excitation \cite{Woerle21}, and theoretical 
results that incorporate Auger recombination and impact ionization indicate that fast 
thermalization processes limit the temperature difference between electrons and holes to 400 K 
\cite{Chin-Yi2019}, negligible compared to our high electronic temperatures. In general, the 
effect of nonequilibrated electrons and holes should have little effect on the overall 
distribution function, if $k_{\rm B}T_{\rm e} \gg E_{\rm gap}$. Here, the corresponding laser-excited electron 
temperature is between $1.27–1.50$ eV, which is almost three times the LDA electronic 
band gap of 0.56 eV.
Moreover, theoretical works suggest energy absorption times to $4-8$ fs 
for $5–10$ fs pulse widths, respectively \cite{Yamada2025}. In agreement, experiments using a 
sub-5-fs light pulse indicate a carrier thermalization time of around $4$ fs and a electron-hole 
gas equilibrates with phonons on a timescale of $58$ fs \cite{Woerle21}. We note that smearing 
out the excitation over $50$ to $100$ fs, resembling a temporal Gaussian excitation with FWHM 
of $50$ fs, has been found to lead to the same crystal response despite a temporal offset 
compared to instantaneous excitation \cite{KEMPKES2021}, which justifies our approximation.

The electronic system is described in the 
microcanonical ensemble, in which the electronic entropy stays constant after the pulse, 
allowing the electronic temperature to change. The results in this paper capture the 
nonthermal contribution caused by highly excited electrons. Equilibration processes between 
electrons and phonons through incoherent electron-phonon energy transfer \cite{Gilberto, 
Sadavisam} are not considered here explicitly. Despite significant progress 
\cite{Sadavisam,Ogitsu2018,Xu2021,Jo2022,Caruso} in accurately describing incoherent 
electron-phonon coupling in computational models, it remains incomplete due to computational 
expense and a missing {\it ab initio} description far from equilibrium. Somewhat simpler 
theories are often used that involve the crucial electron-phonon coupling time at zero 
excitation density $\tau_{\rm 0}$. In silicon reported values range from $\sim 60$ fs 
\cite{Woerle21} to $\sim 115$ fs \cite{Tanimura19} to the often used $240$ fs 
\cite{Sjodin1998} (for electrons $1-2$ eV above the band edge). 
Previous research indicated that electron-phonon interactions in Si increase 
above $17500$ K due to direct nonthermal ion acceleration \cite{Medvedev2023}, which cause a 
decrease in the electron-phonon coupling time. However, our studied excitations are below 
$17500$ K and we intend to stop the ionic motion, so this does not affect our estimate. By 
contrast, there are reports that electron-phonon relaxation should be slower at high excitation 
density due to increasing electronic screening \cite{Sjodin1998,Venkat2022,Klein2021} which 
was reinforced by recent experiments \cite{Swain2025}. We assume the effect of electron screening 
to be dominant.
\begin{figure}[t]
  \centering
    \includegraphics[width=0.8\textwidth]{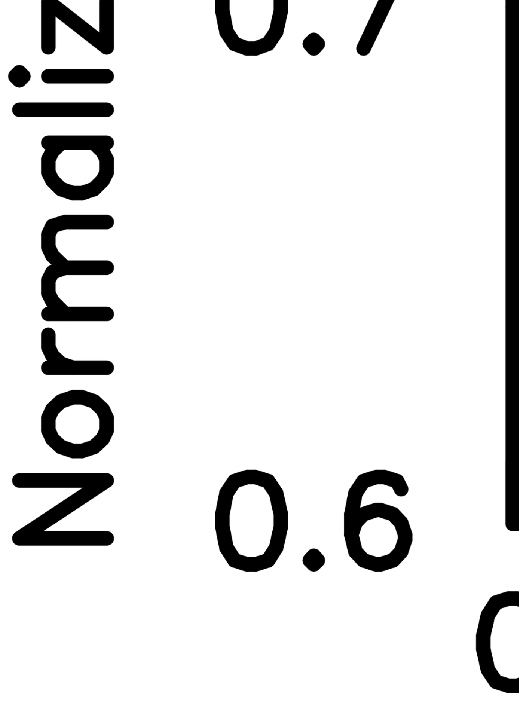}
\caption{\textbf{Time-dependent Bragg intensities:} Averaged time-dependent ($111$) Bragg 
intensity for a two-pulse excitation to initially $T_{\rm e} = 14 842$ K and $T_{\rm e} 
= 17 251$ K after $t=126$ fs (blue), and single-pulse excitations to $T_{\rm e} = 14 842$ K 
(green) and $T_{\rm e} = 17 368$ K (orange). Vertical widths show standard deviations. 
(Inset) Mean effective interatomic potential along acoustic phonon directions 
(see subsection Time-dependent Bragg peak intensities) for the equilibrium structure at $T=315$ K (black solid line) and 
after excitation pulses.}
\label{fig:bragg}
\end{figure}
In almost all studies of electron-phonon equilibration after an intense femtosecond-laser 
pulse the results from Sjodin et al. \cite{Sjodin1998} or modifications are used to include this 
screening effect on the electron-phonon coupling strength \cite{Venkat2022,Klein2021}. 
Applying it to our lowest excitation to $T_{\rm e} = 14 842$ K ($S_{\rm e} = 5.05$ mHa K$^{-1}$), 
this would result in an electron-phonon relaxation time ranging between $32.5$ ps and $130$ 
ps using $\tau_{\mathrm{0}} = 60$ fs or $\tau_{\mathrm{0}} = 240$ fs, respectively, which is much 
longer than our simulation time. Though this is a simple estimate, it seems currently the best way 
to estimate the electronic screening effect \cite{Swain2025} on electron-phonon interactions and is 
commonly used by the community \cite{Venkat2022,Klein2021}. 
In addition, results from density-functional-perturbation-theory (DFPT) found a very strong 
and fast coupling to low-$q$ optical phonons \cite{Park2020,Sadavisam,Gunella2016}. Given 
the complexity of detailed electron-phonon simulations, we estimated an upper bound on the 
impact of fast heated optical phonons via velocity rescaling 
\cite{Toton_2010,Medvedev15,Plettenberg2023,Ihm2024}. We did not find a substantial influence 
on the pausing mechanism within our simulation time, and found that acoustic phonons could 
even end up being cooled slightly. Further details can be found in Methods: Simulation details. 

The computations were performed on a supercell that contains $N=1000$ silicon atoms near 
room temperature. All quantities shown in this paper, except the PhDOS, are averaged 
over ten independently initialized runs because of the thermal ($315$ K) fluctuations. The 
width of the data shown indicates the standard deviations of the averages. In a previous 
study, the analysis of the energy flow in silicon 
\cite{Zier_PRL} shows that for excitations close to but above the nonthermal melting 
threshold, the crystalline structure disorders predominantly in the direction of transverse 
acoustic phonons near the $\Gamma$-point. Based on that knowledge, we designed a 
double-pulse excitation scheme (Fig.\ \ref{fig:scheme}) that can extract energy from 
particular phonon modes, and in contrast to a single excitation to the same excitation 
level, the destabilization of the crystalline structure within the first picosecond is 
prevented. The analytic solution for an ensemble of harmonic uncoupled phonon modes even 
suggests that such a scheme could stop the phonon motion altogether (Supplementary Note 1), 
which shows the principal idea of our approach. 

\subsection{Timed double-pulse excitation scheme}

\begin{figure}[ht]
  \centering
    \includegraphics[width=1.0\textwidth]{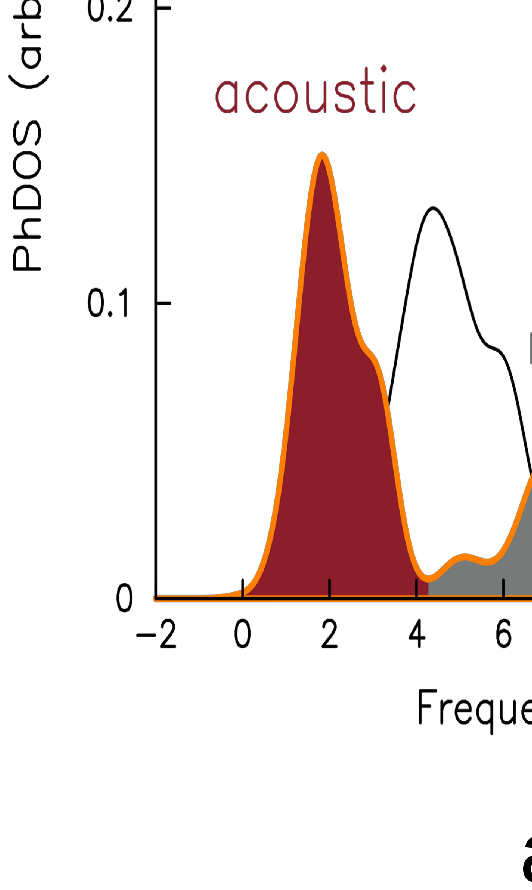}
\caption{\textbf{Ionic temperatures:} a) Phonon density of states of the ground state (GS (black)) and 
for $T_{\rm e} = 17 368$ K (orange), divided into three ranges, namely acoustic (red area), mixed (gray area), and optical (purple area). b) Ionic temperature evolution for the 
double-pulse (blue) and single-pulse excitations (green and orange). c)-d) Ionic temperature evolution and 
decomposition into the three phonon ranges for c) the double-pulse excitation 
(total in blue), and (d) the single-pulse excitation to $T_{\rm e} = 17 368$ K (total in orange).
\label{fig:temp}}
\end{figure}

In more detail, our excitation scheme consists of a first excitation with an absorbed 
fluence\cite{PRX2013} around $10.5$ mJ cm$^{-2}$ , inducing an electronic entropy of 
$S_{\rm e} = 5.05$ mHa K$^{-1}$ with a corresponding electronic temperature of $T_{\rm e} = 14 842$ K. 
This fluence is below the laser-induced nonthermal melting threshold 
\cite{Zier14}, meaning that no irreversible structural changes are induced by the laser, or 
in other words, criteria for melting like the Lindemann criterion are not reached. 
In a previous work \cite{Zier14} we found that the nonthermal melting threshold for silicon lies between 
$16736$ K and $17051$ K or $53$ and $54$ mHa, respectively. In bond-softening materials 
\cite{Stampfli90} such as silicon, a coherent oscillatory ionic motion will be induced, called 
thermal phonon squeezing that was already experimentally observed 
in Ge \cite{Trigo2013} and Bi \cite{Johnson09}.
It was found that this phenomenon is the precursor to 
nonthermal melting \cite{PRX2013} -- in particular, the same transverse acoustic phonons 
that cause nonthermal melting are predominant. Due to the nature of thermal phonon squeezing, 
the induced oscillations in system quantities are frequency-doubled \cite{PRX2013} from 
the underlying phonon frequency. For instance, the first maximum in the mean-square 
displacement ($msd$) (see Methods: Mean-square atomic displacement) and first minimum of the Bragg intensity (Fig.\ 
\ref{fig:bragg}) are reached after a half period, at $t_2=126$ fs. By choosing this $t_2$ 
as the delay time of the second pulse, we are able to specifically address these phonons 
that predominantly cause nonthermal melting. Therefore, time $t_2 = 1/(4\nu)$ 
coincides with a phonon frequency of $\nu \approx 2$ THz, matching our analytic 
analysis (Supplementary Note 1). This frequency is the first peak in the phonon density 
of states (PhDOS) of the excited state (Fig.\ \ref{fig:temp}a)), corresponding to purely 
acoustic phonons. In addition, we chose 
its corresponding induced-electronic entropy to be $S_{\rm e} = 6.14$ mHa K$^{-1}$, 
which is above the ultrafast melting threshold and yet has no lattice instabilities, i.e., 
repulsive phonon directions \cite{Zier_PRL}. However, the initial kinetic energy is 
large enough to overcome the laser-softened interatomic potential barriers. Those 
parameters maximize the pausing effect on laser-induced disordering (see Methods: Mean-square atomic displacement). 
We note that a single-pulse excitation to $S_{\rm e} = 6.14$ mHa K$^{-1}$ corresponds to an 
electronic temperature of $T_{\rm e} = 17 368$ K.
We like to note that our theoretically proposed excitation scheme could 
be adapted to experimental time-resolved pump-pump-probe diffraction measurements using 
either X-Rays or ultrafast electrons. In order to meet our assumptions we suggest excitations 
around the electronic band gap by using Gaussian ultrashort-laser pulses with a temporal FWHM 
of around $10$ fs to capture any ionic motion accurately, with an intensity in the linear 
regime for absorption on the order of $10^{12}$ W cm$^{-2}$. Using the absorbed energy, we 
estimated the fluence of both pulses to be $10.5$ mJ cm$^{-2}$  and $14.5$ mJ cm$^{-2}$ for the 
reported laser-deposited electronic entropy of $S_{\rm e} = 5.05$ mHa K$^{-1}$ and $6.14$ mHa K$^{-1}$, 
respectively. Here, we assumed a penetration depth of 10 nm for near-ultraviolet light 
\cite{Speiser2020,Korpusenko_2023}. 

\subsection{Time-dependent Bragg peak intensities}

The crystalline disordering is accessible in experiment by measuring time-dependent Bragg 
peak intensities. Our computed intensity (Supplementary Note 2 and 4) after the double-pulse excitation 
scheme shows a clearly different time-evolution than the single-pulse excitation with the 
same electronic entropy (Fig.\ \ref{fig:bragg}). After $1$ ps both intensities differ 
by roughly $25$\%. Whereas the system after the single pulse shows a monotonic decrease of 
the Bragg intensity for $t > 400$ fs, indicating a nonthermal disordering process, 
the double-pulse intensity remains within any stabilization limit, e.g. Lindemann criterion.
We note that this main behavior of the intensities does not change in the presence of an 
effective incoherent electron-phonon coupling modelled by ultrafast optical phonon heating 
(Supplementary Methods). The phase after the second pulse, while diamond-like in short-range order, 
is also different from the equilibrium structure at $315$ K, which can be seen by the drop of 
$\sim 8$\% in the first $200$ to $300$ fs. The energy difference to an equilibrated system at 
the same conditions for $T_{\rm e}, T_{\rm I}$ is $0.086 \ \mathrm{eV} \mathrm{atom}^{-1}$ 
(See Methods: Simulation details), comparable to differences between different phases of Si 
\cite{Malone2012,Wippermann2016}, and between 
crystalline and amorphous Si \cite{Kail,Drabold}, which indicates a resemblance to a 
solid-solid phase transition. Moreover, the pair-correlation function (Supplementary Figure 9) of the system indicates that the short-range order (below $5$ \AA) in this state is 
comparable to the equilibrium structure, but the long-range order is washed out and loses its 
fine structure (Supplementary Note 3). Similar trends can be seen in the experimentally accessible structure function S(q) (Supplementary Figure 10).

We find this behavior is explained by the potential-energy surface (PES), 
as computated for different electronic temperatures along the transverse acoustic mode at 
$q = (0.2, 0, 0) \ 2\pi/ a$, with $a$ the lattice parameter (Methods: Simulation details), which mainly 
drives the disordering process \cite{Zier_PRL}. For every timestep $t$ of our MD simulation 
we projected the force vector $\vec{F}(t)$ and the displacement vector $\vec{r}(t) - \vec{r}_{\mathrm{0}}$
from  the ideal positions in the diamond-like crystal, onto the eigenvector $\vec{e}_j$ of the 
$j$-th phonon of the excited state ($T_{\rm e} = 17 368$ K). This enabled us to connect the force 
$F$ and the ionic displacement $u$ in the direction of the phonon mode for every timestep. 
In summary, we used the ions to probe the effective force in the direction of the phonon. In 
a next step we fitted those data sets of $F(u)$ to a third-order polynomial, which is then 
integrated to obtain the potential energy surface in this direction. 
Our results allow direct 
insights into our MD simulations and go beyond previous schematic, static or fitted models 
\cite{PRX2013,Zier_PRL,Bauerhenne2020}. In order to obtain results that do not depend on the 
initialized thermal conditions, we averaged over all $12$ equivalent phonon modes and the 
ten independent simulation runs mentioned above (Supplementary Note 5). The inset of Fig.\
\ref{fig:bragg} summarizes those results by showing the mean values of the potential energy 
for all excitations used and the equilibrium structure at $315$ K. Whereas the single-pulse 
excitation above the ultrafast melting threshold has one of its barriers lowered, the 
potential energy of the double-pulse excitation retains higher and symmetric potential barriers 
($\sim 0.04$ eV), which in the end preserves the short-range order of the crystal and is 
responsible for the pausing of nonthermal melting. As a result, the atomic 
motion is frozen, which means that the atoms are static around their positions $\vec{r}(t_2)$ 
(Supplementary Note 7).

\subsection{Ionic-temperature}

Moreover, the time evolution of the ionic temperature, computed via the equipartition relation 
$\langle E_{\rm kin} \rangle = \frac{3}{2} \left( N - 1 \right)k_{\rm B} T_{\rm I}$ (Boltzmann 
constant $k_{\rm B}$), shows that the double-pulse excitation scheme has an energy-extracting 
effect, which we also demonstrated in the harmonic approximation (Supplementary Note 1). For ions 
with low kinetic energies that are near their turning points, the change 
in potential energy surface due to the second pulse prevents them from gaining kinetic energy 
again, which cools the system for longer times (Fig.\ \ref{fig:temp}b)). For all three 
studied pulse schemes, the ionic temperature drops within $\sim 200$ fs. However, only after 
the double-pulse excitation does it remain constant at $\sim 215$ K despite the much higher 
electronic temperature, whereas for both single pulses the temperature increases again (Fig.\ 
\ref{fig:temp}b)). This temperature increase is attributed to ions gaining again kinetic 
energy after the turning point in an attractive potential after an excitation below the 
nonthermal melting threshold or for excitations above that threshold by ions overcoming the 
potential barriers softened by the laser. In the latter case, ions are accelerated away from
their equilibrium positions causing disorder. 
We note that Fig. \ref{fig:temp}) shows the behavior of the ionic temperature in the presence of only coherent energy exchange. Supplementary Figures $4$, $5$, and $6$ show how the ionic temperature increases in the presence of electron-phonon equilibration processes.

Figure \ref{fig:temp}a) shows the phonon density of states (PhDOS) of the ground state as well as 
of the excited state ($T_{\rm e} = 17 368$ K). The PhDOS \cite{PRX2013,Zier15} is divided into three 
ranges (acoustic, mixed, and optical) to identify which phonons are mostly affected by the 
changes in the excitation scheme. We computed partial kinetic energies by projecting the 
ionic velocities on the phonon eigenvectors of the excited state ($T_{\rm e} = 17 368$ K) within 
the corresponding ranges. Using again the equipartition relation, we were able to calculate 
the corresponding partial ionic temperatures \ref{fig:temp}c). The fact that different phonon ranges do not 
share a common temperature during our simulation was seen previously \cite{PRX2013} but the 
large difference of around $100$ K is remarkable. The slopes of the temperature curves suggest 
a phonon-phonon thermalization time of about $6$ ps (much faster than predicted phonon 
equilibration times of several nanoseconds at lower energy \cite{Zhao2008}), suggesting this 
is the timescale for persistence of the controlled phase. Note that this finding is resilient 
to fast optical phonon heating by electron-phonon coupling within the first ps (Supplementary Method) 
and that all phonon-phonon interactions are considered in our approach. 
\begin{figure}[ht]
  \centering
    \includegraphics[width=1.0\textwidth]{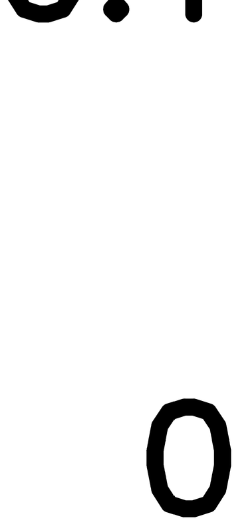}
\caption{\textbf{Electronic band gap:} Time evolution of the Kohn-Sham band gap.
The band gap decrease induced by a single pulse to $T_{\rm e}= 17 368$ K (orange) 
can be paused by using two pulses instead (blue), similar to a single pulse to $T_{\rm e} = 14 842$ K 
(green). (Inset) At $1$ ps after the excitation with two pulses, ground-state 
characteristics (grey, covered by green) are retained near the band gap.}
\label{fig:bandgap}
\end{figure}
In the single-pulse case (Fig.\ \ref{fig:temp}d)) a common temperature of the 
phonon subsystem could be reached within the simulation time, mainly caused by 
disordering over-the-barrier motion (Fig.\ \ref{fig:temp}d)). In contrast, our 
proposed pausing mechanism prevents over-the-barrier ionic motions and therefore 
reheating of acoustic phonon modes. Instead, we created a highly 
excited solid system with cooled phonons. 

\subsection{Electronic density of states - Band gap}

Besides the described structural behavior we also found different but not exotic electronic 
properties, when compared to both the ground state and the laser-disordered state (Fig.\ 
\ref{fig:bandgap}). In the latter case, the band gap disappeared within $600$ fs after the 
excitation, indicating a semiconductor-to-metal transition \cite{Silvestrelli96,McMillan2005, 
Medvedev15}. However, after the double pulse, the structure remains semiconducting in the 
time studied. Moreover, the density of states (DOS) characteristics of the ground state near 
the Fermi energy are preserved. For example, the ``V'' shape of the DOS and the small 
oscillations are comparable to the ground state. However, differences from the ground state 
are noticeable, e.g. the oscillations are less pronounced and the conduction bands are 
shifted down slightly, which decreases the band gap by nearly $30$\%.

\section*{Conclusions}
In summary, we show a paradigm of control using laser-induced bond-softening effects. This 
non-typical control mechanism allows us to direct the ionic motion on a highly excited 
potential energy surface by light and is general to bond softening materials, as suggested 
by our analytic study of a single laser-excited phonon mode. Moreover, our mechanism is 
resilient to ultrafast optical phonon heating by incoherent electron-phonon coupling within 
our simulation time (Supplementary Method), which makes it interesting for experimental realization. 
Moreover, our results indicate that a pausing effect is still significant 
up to an electronic temperature of 63 mHa (19894 K) induced by the second pulse (Supplementary 
Figure 16). So, the effect should be observable at least over an electronic temperature range of 
2500 K. Several physical effects impose an upper limit. (1) With our proposed excitation scheme 
crystal symmetry loss by nonthermal melting is only delayed compared to a single-pulse excitation, 
but at higher electronic temperatures there are increased decoherences and anharmonicities which 
will hinder the stopping of ionic motions by the mechanism we described. (2) For much larger 
electronic excitations, lattice instabilities are induced for multiple phonon directions and the 
potential energy surface may also become repulsive, so redirection or pausing of atoms seems 
unlikely. (3) If the first pulse is already well above the nonthermal melting threshold the 
structure will melt anyway, even if a second pulse is applied.

This proof-of-principle of the proposed excitation mechanism also applies to other 
systems with weak electron-phonon interaction like InSb \cite{McCombe1969}. To realize 
pausing in other materials as well, we suggest that the optimal timing of the second pulse 
can be directly obtained by measuring the first maximum in the mean-square ionic displacement 
and/or the first minimum in the corresponding oscillation in the Bragg peak intensity caused 
by the first laser pulse (Fig.\ \ref{fig:bragg}). In materials with lower symmetry and/or more 
degrees of freedom, it might be possible to generate stable or metastable 
phases by solid-to-solid phase transitions \cite{Wippermann2016}. Moreover, the application 
to more complex systems than silicon might have the 
potential to create novel phases in both structure and electronic properties. For such systems, 
optimal control approaches \cite{Kohler2011,Katsuki2018,Siegrist2019,Qasim2021,Rupprecht2022} 
could be used to obtain the number of pulses in a train and their polarization, phase and shape 
for maximal efficiency. In addition, our suggested pausing approach could enable time-resolved 
experiments to obtain reliable equilibration times between electrons and phonons, and therefore 
electron-phonon coupling coefficients by measuring the temporal deviation, e.g., in the 
time-evolution of the Bragg peak intensity. In more detail, measuring the slope of a Bragg peak 
decay after a single pulse vs. after our proposed scheme would allow a direct measurement of 
the incoherent heat transfer to the phonons. In turn, those observed values can then be used 
to improve theoretical descriptions of thermal contributions to laser-induced structural 
changes \cite{Ogitsu2018}. All in all, our proposed excitation scheme is a paradigm of 
material control out of equilibrium and could lead to insights in light-matter interactions 
and improve Floquet engineering as well as electron-phonon interaction approaches. 

\section*{Methods}

\subsection{Simulation details}

\begin{figure}[t]
  \centering
\includegraphics[width=0.8\textwidth]{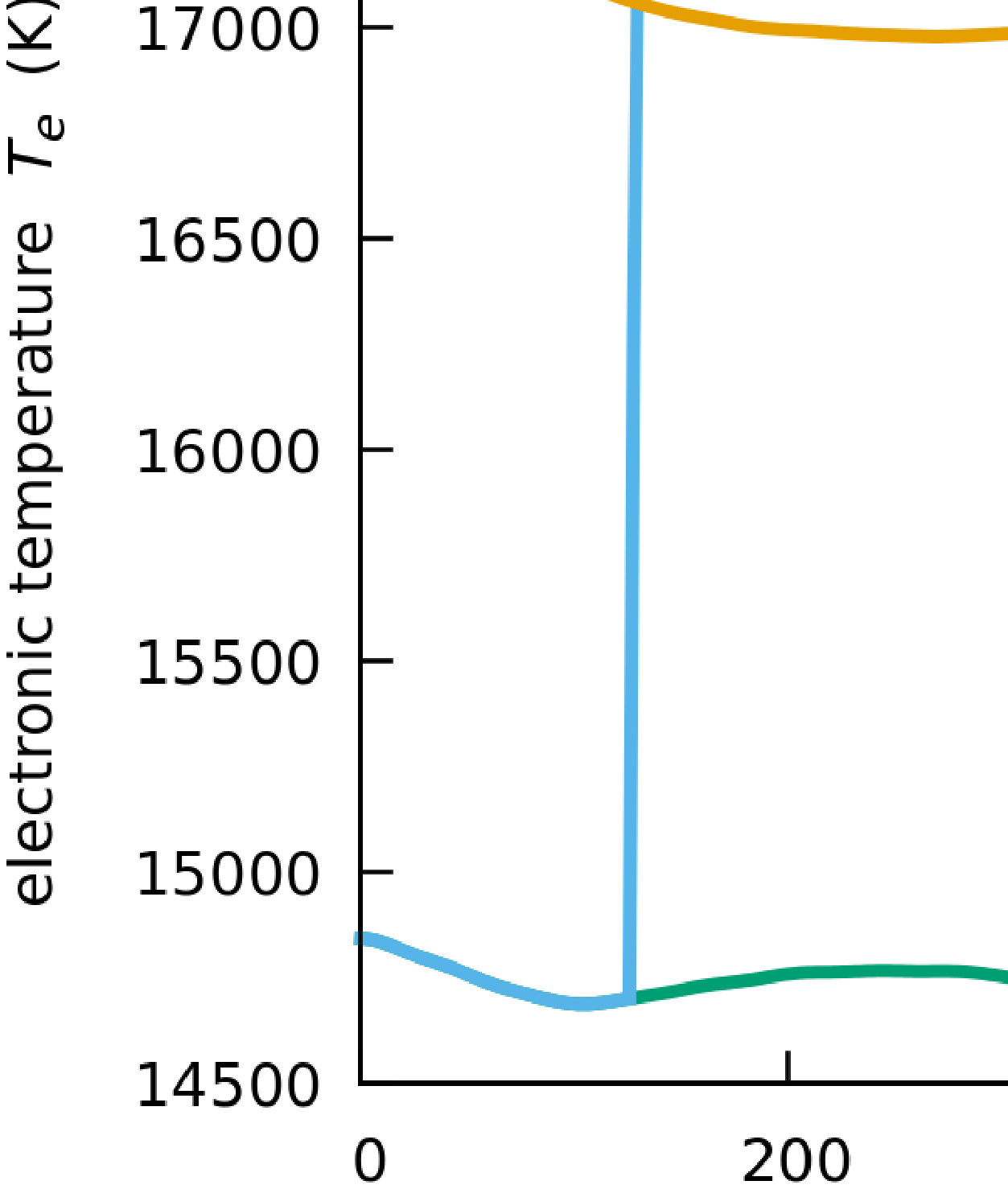}%
\caption{\label{fig:el_temp}\textbf{Electronic temperature:} Time-evolution of the electronic temperature $T_{\rm e}$ for the 
 single pulse excitations to $T_{\rm e}=14 842$ K (green) and $T_e=17368 $ K (orange), as well as for 
 the double-pulse excitation to $T_{\rm e}= 17 251$ K (blue).}
\end{figure}

The finite-temperature DFT code CHIVES uses atom-centered Gaussian 
basis sets and norm-conserving pseudopotentials \cite{PRX2013,Zier15,Zier_PRL} to 
describe the electronic system. The exchange and correlation energy is calculated 
in the local density approximation \cite{LDA}. The excitation by a femtosecond-laser 
pulse can be approximated in CHIVES by running simulations at either constant electronic 
temperature or entropy \cite{PRX2013,Grigoryan2014,Zier2017}. In the mode of constant 
electronic temperature, which can be realized by a connection to an infinite heat bath, 
the temperature is the input value for the excitation strength. The corresponding Fermi 
distribution is used to determine the electron occupations. In such case, the proper 
description of the electronic system is the canonical ensemble ($N,V,T$), in which the 
free energy $F = E -T_{\rm e}S_{\rm e}$ is a constant. This simulation mode is a good description of 
very intense excitations where small variations in $T_e$ are negligible and/or for very 
short simulation times of a few hundred fs after the excitation in which the electronic 
temperature does not have time to change dramatically. For simulations with constant 
entropy the electronic system is described by the microcanonical ensemble ($N,V,E$). In 
this case the total energy $E$ is constant but the electronic temperature $T_e$ can 
change. This simulation mode is a more physical description for longer simulation times 
and moderate excitations. Here, we performed constant entropy simulations. Figure 
\ref{fig:el_temp} shows the time-evolution of the electronic temperatures for the three 
different excitations described in the main text. In particular the excitation to 
$S_{\rm e} = 6.14 $ mHa K$^{-1}$ ($T_{\rm e} = 17 368$ K) in a single pulse, after which 
the crystal melts nonthermally, shows a significant change in the electronic temperature 
within our simulation time. Using the proposed double-pulse excitation prevents energy 
transfer from the highly excited electronic system, indicated by a more or less constant 
value of $T_{\rm e}$.  
In addition, we investigated the potential-energy difference of this laser-induced 
transition to a metastable phase after the second pulse. To make this comparison, 
we need two calculations at common $T_{\rm I}$ and $T_{\rm e}$. We use the energies of the 
ionic snapshots of the system after the second pulse at $1$ ps, with $T_{\rm e} = 315$ K 
and $T_{\rm I} \approx 215$ K (the average, as shown in Fig. 3c), and calculate the 
total energy difference from a reference system close to the equilibrium structure, 
which was initialized at $T_{\rm I} = 215$ K and $T_{\rm e} = 315$ K. The result 
$ \Delta E = 0.086 \ \mathrm{eV/atom}$ is in the range of energy differences 
between solid phases in silicon \cite{Malone2012}. 

We use a supercell consisting of 
$5\times5\times5$ conventional unit cells of $8$ atoms each. The unit cell itself 
has an internal lattice parameter of $5.3987$ {\AA}. Due to the large supercell 
it is enough to perform simulations in the $\Gamma$-point approximation 
to obtain convergence. The atomic displacements and velocities are distributed 
according to a Maxwell distribution with $k_{\rm B}T_I = 1\ \mathrm{mHa}, T_I\approx 
315$ K by using true random numbers \cite{random} and the procedure explained in 
\cite{PRX2013}. We generated $N_{\rm run} = 10$ molecular-dynamics simulation runs 
with independent initial ionic conditions. Their trajectories are averaged to 
obtain the presented quantities, e.g., the mean-square atomic displacement. The 
time-evolution of the atomic system is modelled by a velocity-Verlet algorithm 
using a timestep of $2$ fs, which was already tested in previous studies
\cite{PRX2013,Zier2017}. For the electronic density of states (EDOS) we used 
the energy eigenvalues of the system, which were smeared by Gaussians with a 
FWHM of $0.2$ eV. For all excitations the ionic configuration of the last 
timestep of our simulation (at $1$ ps) was considered, respectively. For the phonon 
density of states (PhDOS) we diagonalized the dynamical matrix, which was obtained 
by finite differences with $\Delta R = 0.001 a_0 = 0.000529$ {\AA} for the system 
at $T_I = 0$ K, with no thermal displacements. The phonon eigenvalues are 
then also smeared by Gaussians with a FWHM of $1$ THz. The eigenvectors $\vec{e}_j$ 
of the phonons were used to compute particular phonon contributions to the ionic 
temperature (see main text) or to compute the effective atomic potential. 
We computed the phonon eigenvectors and values for $S_{\rm e} = 5.05$ mHa K$^{-1}$ 
($T_{\rm e} = 14 842$ K) and $S_{\rm e} = 6.14$ mHa K$^{-1}$ ($T_{\rm e} = 17 368$ K) 
and found very small 
differences and basically no different results for the projected quantities, i.e. ionic 
temperature or interatomic forces. In addition, we compared those results to the phonon 
eigenvectors in the canonical ensemble at comparable excitation densities 
($T_{\rm e} = 17 368$ K), which gave basically same results. Therefore, we show in the main text the 
results projected onto only the eigenvectors corresponding to $S_e = 6.14$ mHa K$^{-1}$. 

\subsection{Mean-square atomic displacement}

As mentioned in the main text, our mechanism is determined by three parameters: 
the electronic entropy or temperature induced by the first pulse, the timing $t_2$ of the 
second pulse, and its induced electronic entropy or temperature above the ultrafast melting 
threshold. In order to find a parameter set that shows a large effect on 
\begin{figure}[t]
  \centering
\includegraphics[width=0.8\textwidth]{Zier_fig6.ps}
\caption{\label{fig:ms} \textbf{Mean-square atomic displacement:} Time-dependence of the mean-square atomic displacement for 
a single pulse excitation to $T_{\rm e}= 14 842$ K ($S_{\rm e} = 5.05$ mHa K$^{-1}$), our proposed 
double-pulse excitation to a final $S_{\rm e} = 6.14$ mHa K$^{-1}$ ($T_e = 17 251$ K), and a 
single pulse excitation to $S_{\rm e} = 6.14$ mHa K$^{-1}$ ($T_{\rm e} = 17 368$ K). The vertical 
thickness of the lines shows one standard deviation above and below the mean.}
\end{figure}
the laser-induced disordering, we performed simulations for various sets of electronic 
temperatures and computed the mean-square atomic displacement ($msd$) (Fig.\ \ref{fig:ms}), which 
is calculated using the equation 
\begin{equation}
msd(t) = \sum_{j}^{N_{\rm run}} \frac{ \sum_{i}^{N}\left(\vec{r}^{j}_{i}(t) - 
\vec{r}_{i}^{0}\right)^2}{N} \frac{1}{N_{\rm run}}\; ,
\label{MSD_r0}
\end{equation}
where $N_{\rm run}=10$ the number of independent runs, $N=1000$ the number of atoms, 
$\vec{r}_{i}(t)$ is the position vector of atom $i$ in run $j$ at time $t$, and 
$\vec{r}_{i}^{0}$ is the ideal silicon structure at $T=0$ K. Some of the results are shown 
in Fig.\ \ref{fig:ms_comp}. A small and un-changing value of the $msd$ indicates a 
stopped laser-induced non-thermal, ultrafast melting process.   
\begin{figure}[ht!]
  \centering
\includegraphics[width=0.9\textwidth]{Zier_fig7.ps}
\caption{\label{fig:ms_comp} \textbf{Different excitation parameters:} Mean-square displacement data for $t_{\rm 2} = 100, 122, 
126$, and $160$ fs, and $T_{\rm e} = 17 684$ K. As a reference the mean-square displacement of $t_{\rm 2} 
=126$ and $T_{\rm e}=17368 $ K (light blue) and for the single pulse excitation with 
$T_{\rm e} = 17 684$ K is shown (light green).}
\end{figure}
For our parameters presented in the main text, the $msd$ increases after the first pulse 
but remains constant after the second one. For changes in $t_2$ or the induced 
electronic temperature of the second pulse, we observe that the $msd$ instead increases 
significantly after a few hundreds of fs following the double-pulse excitation.
The slope at $1$ ps in these other cases, as the $msd$ continues to grow, implies 
that our suggested mechanism is not used effectively for those other parameters. 
In such a case the prevention of the laser-induced disordering is achieved only for a 
shorter timescale than by the parameters used in the main text. The $msd$ data shows 
characteristics similar to the Bragg peak intensities, e.g., the time for the 
first local maximum of the $msd$ after the first pulse matches the time of the 
first local minimum of the Bragg peak intensity.

\section*{Acknowledgments}
Computations were performed at the Lichtenberg High Performance Computer of the Technical University Darmstadt and the Pinnacles cluster at UC Merced, supported by 
National Science Foundation Award OAC-2019144. T.Z. was supported by the Deutsche 
Forschungsgemeinschaft through the project ZI 1858/1-1. D.A.S. was supported by the 
U.S. Department of Energy, National Nuclear Security Administration, Minority 
Serving Institution Partnership Program, under Award DE-NA0003984.
T.Z. and D.A.S. were also supported by the Multi-Campus Research Programs and Initiatives of the University of California, Grant Number M23PR5854. M.E.G was 
supported by the Deutsche Forschungsgemeinschaft through the project GA 465/27-1  
and the Federal Ministry of Education and Research (BMBF) through project DIQTOK.
\section*{Author Information} 
Correspondence should be addressed to T.Z.\ (tzier2@ucmerced.edu) or D.A.S.\ (dstrubbe@ucmerced.edu).

\section*{Competing Interest}
All authors declare no competing interests.

\section*{Data availability statement}
Raw datasets generated during the current study for Figures 2, 3, and 4 are available at Zenodo - EU Open Research repository under https://doi.org/10.5281/zenodo.15947394  \\
Additional data of this study are available from the corresponding author upon reasonable request.

\section*{Code availability}
The used code CHIVES to generate the findings of this study is available from the corresponding author upon request.

\section*{Author contribution}
M.E.G and E.S.Z. initially discussed the presented idea. T. Z. performed the MD simulations, analyzed the data, and prepared the figures. T. Z. and D. A. S. revisited the data and re-evaluated the results. T. Z. implemented the effective electron-phonon coupling and wrote the first version of the manuscript. All authors discussed the results 
and worked on the manuscript.

\section*{Additional information}
\textbf{Supplementary information} The online version contains supplementary material
available for double-pulse excitation of one phonon-mode in the harmonic potential, additional Bragg peak intensities, effects of optical phonon heating by electron-phonon coupling, pair-correlation functions, structure functions, effective potential energy surfaces, partial ionic temperatures, atomic quasi-stationary state after the second pulse, and estimated temperature range of effect observability.

\section*{References}
\bibliography{apssamp}

\end{document}


\preprint{APS/123-QED}

\title{\textit{Supplemental Material for:} Pausing ultrafast melting by timed multiple femtosecond-laser pulses\\
}

\author{Tobias Zier}
\email{tzier2@ucmerced.edu}
\affiliation{Department of Physics, University of California Merced, Merced, CA 95343}%
\affiliation{%
 Theoretical Physics, University of Kassel, Heinrich-Plett-Str.\ 40, 34132 Kassel, Germany
}%

\author{Eeuwe S.\ Zijlstra}%
\affiliation{%
 Theoretical Physics, University of Kassel, Heinrich-Plett-Str.\ 40, 34132 Kassel, Germany
}%
\affiliation{%
Center for Interdisciplinary Nanostructure Science and Technology 
(CINSaT), Heinrich-Plett-Str.\ 40, 34132 Kassel, Germany
}%

\author{Martin E. Garcia}
\affiliation{%
 Theoretical Physics, University of Kassel, Heinrich-Plett-Str.\ 40, 34132 Kassel, Germany
}%
\affiliation{%
Center for Interdisciplinary Nanostructure Science and Technology 
(CINSaT), Heinrich-Plett-Str.\ 40, 34132 Kassel, Germany
}%

\author{David A. Strubbe}
\affiliation{Department of Physics, University of California Merced, Merced, CA 95343}%


\date{\today}

\maketitle


\section*{\textbf{Supplementary Note 1:} Double-pulse excitation of one phonon-mode in the harmonic potential}\label{sec:harmonic}

\begin{figure*}[h]
\includegraphics[width=0.8\textwidth]{sketch_harm.ps}%
\caption{\label{fig:sketch_harm} Sketch of the double-pulse excitation in the harmonic potential. (Left) 
The atomic displacements in the ground state are consistent with the potential. (Middle) The first 
laser pulse induces a potential change so that the phonon frequency changes from $\omega_{0}$ to 
$\omega_{1}$. On average, the atoms are now moving away from their initial positions. (Right) 
At some later time $t_2$, a second pulse is applied that induces an additional potential change, which causes the phonon frequency to change to $\omega_{2}$.}
\end{figure*}

As a proof of concept we solved analytically the double-pulse excitation for an ensemble of 
uncoupled phonon modes, all with the same frequency in the harmonic approximation. Suppose  
in the ground state (Suppl. Fig. \ref{fig:sketch_harm}) all those modes have frequency $\omega_{0}$ as determined by the potential energy surface (PES).
For a system that was thermalized at $T_I$ 
before the laser pulse, the initial displacements $u_{0}$ and velocities $v_{0}$ are given 
by the corresponding Maxwell distributions 
\begin{align}
f_{u}(T_I, u_{0})&= \sqrt{\frac{m\omega_{0}^{2}}{2\pi k_{\rm B}T_{I}}}\exp\left({-\frac{m\omega_{0}^{2}}{2k_{\rm B}T_{I}}u_{0}^{2}}\right) 
\label{eq:f(u)}
\end{align}
\begin{align}
f_{v}(T_I,v_{0})&= \sqrt{\frac{m}{2\pi k_{\rm B}T_{I}}}\exp\left({-\frac{m}{2k_{\rm B}T_{I}}v_{0}^{2}}\right) \; ,
\label{eq:f(v)}
\end{align} 
where $m$ is the atomic mass and $k_{\rm B}$ is the Boltzmann constant. Applying now the 
first pulse induces a change in the potential energy (Suppl. Fig. \ref{fig:sketch_harm}, middle). 
Accordingly, the frequencies of the phonon eigenmodes in the new potential become $\omega_{1}$. 
Here, we consider an instantaneous potential change in which the statistical distribution from the ground 
state persists, before a new equilibration is established. The time-dependent atomic displacement 
$u(t)$ and the corresponding velocity $v(t)$ of the phonon modes can be written as 
\begin{align}
\label{eq:u_square}
u(t)&= u_{0}\cos(\omega_1 t)+ \frac{1}{\omega_1}v_{0}\sin(\omega_1 t)
\end{align}
\begin{align}
v(t)&= -u_{0}\omega_1\sin(\omega_1 t)+v_{0}\cos(\omega_1 t) \; ,
\end{align}
where $u_{0}$ is the initial displacement at $t=0$ and $v_0$ is the initial phonon velocity. 

In order to obtain $u^{2}(t)$, we have to integrate over all possible displacements and initial 
velocities from the Maxwell distributions:
\begin{align}
\label{first_puls_msd_a}
\langle u^{2}(t)\rangle &= \int\int du_{0}dv_{0} (u(t))^2 f_{u}(T_I) f_{v}(T_I) \;. 
\end{align} 
Using Supplementary Eq. \ref{eq:u_square} and Supplementary  Eqs. \ref{eq:f(u)} and \ref{eq:f(v)} we get:
\begin{align}
\label{first_puls_msd_b}
\langle u^{2}(t)\rangle &= \int\int du_{0}dv_{0} \sqrt{\frac{m\omega_{0}^{2}}{2\pi k_{\rm B}T_{I}}}\exp\left({-\frac{m\omega_{0}^{2}}{2k_{\rm B}T_{I}}u_{0}^{2}}\right) \sqrt{\frac{m}{2\pi k_{\rm B}T_{I}}}\exp\left({-\frac{m}{2k_{\rm B}T_{I}}v_{0}^{2}}\right) \notag \\
&\left[ u_{0}^{2}\cos^{2}(\omega_1
t)+ \frac{v_{0}^{2}}{\omega_1^{2}}\sin^{2}(\omega_1 t)+2uv\sin(\omega_1
t)\cos(\omega_1 t)\right] \;. 
\end{align}
After sorting the terms we get
\begin{align}
\langle u^{2}(t) \rangle &= \int dv_{0} \left[\int du_{0} \frac{m\omega_{0}}{2\pi k_{\rm B}T_{I}}\cos^{2}(\omega_1 t) u_{0}^{2}\exp\left({-\frac{m\omega_{0}^{2}}{2k_{\rm B}T_{I}}u_{0}^{2}}\right)\right]\exp\left({-\frac{m}{2k_{\rm B}T_{I}}v_{0}^{2}}\right) \notag \\
&+ \int du_{0} \left[ \int dv_{0} \frac{m\omega_{0}}{2\pi k_{\rm B}T_{I}}\frac{\sin^{2}(\omega_1 t)}{\omega_1^{2}}v_{0}^{2}\exp\left({-\frac{m}{2k_{\rm B}T_{I}}v_{0}^{2}}\right)\right]\exp\left({-\frac{m\omega_{0}^{2}}{2k_{\rm B}T_{I}}u_{0}^{2}}\right) \notag \\
&+ \underbrace{\int\int
du_{0}dv_{0} \sqrt{\frac{m^{2}\omega_{0}^{2}}{4\pi^{2}k_{\rm B}^{2}T_{I}^{2}}}
2uv\sin(\omega_1 t)\cos(\omega_1 t)
\exp\left({-\frac{m\omega_{0}^{2}}{2k_{\rm B}T_{I}}u_{0}^{2}}\right)\exp\left({-\frac{m}{2k_{\rm B}T_{I}}v_{0}^{2}}\right)}_{=0,\text{
since} \int dx x e^{-\alpha x^{2}}=0} \;,
\label{eq:msd_harmonic}
\end{align}
in which the third term equals zero in any case. Using Gaussian integral relations for terms $1$ and 
$2$ we get
\begin{align}
\langle u^{2}(t) \rangle &= \cos^{2}(\omega_1 t) \frac{m\omega_{0}}{2\pi k_{\rm B}T_{I}} \frac{\pi}{2\left(\frac{m\omega_{0}^{2}}{2k_{\rm B}T_{I}}\right)^{\frac{3}{2}} \left(\frac{m}{2k_{\rm B}T_{I}}\right)^{\frac{1}{2}}} \notag \\
&+  \sin^{2}(\omega_1 t) \frac{1}{\omega_1^{2}}\frac{m\omega_{0}}{2\pi
k_{\rm B}T_{I}}\frac{\pi}{2\left(\frac{m\omega_{0}^{2}}{2k_{\rm B}T_{I}}\right)^{\frac{1}{2}} \left(\frac{m}{2k_{\rm B}T_{I}}\right)^{\frac{3}{2}}} \; , 
\end{align}
which can be reduced to 
\begin{equation}
\langle u^{2}(t) \rangle  = \cos^{2}(\omega_1
t) \frac{k_{\rm B}T_{I}}{m\omega_{0}^{2}}+\sin^{2}(\omega_1
t)\frac{k_{\rm B}T_{I}}{m\omega_1^{2}} \;.
\label{eq:reult_msd_one_pulse}
\end{equation}
The maximum of the mean-square displacement $\langle u^{2}(t)\rangle$ can be computed by finding when this condition is satisfied:
\begin{align}
\frac{\partial \langle u^{2}(t) \rangle }{\partial t} & = 0 \notag \\
& = \frac{k_{\rm B}T_{I}}{m} \left[-\frac{2 \omega_{1}}{\omega_{0}^{2}}\cos(\omega_{1} t) \sin(\omega_{1} t) 
 + \frac{2 \omega_{1}}{\omega_{1}^{2}} \cos(\omega_{1} t) \sin(\omega_{1} t) \right] \\
\end{align}
Using the relation:
\begin{equation}
\cos(\omega_{1} t) \sin(\omega_{1} t)  = \frac{1}{2} \sin(2\omega_{1} t)
\end{equation}
we get

\begin{align}
& = \frac{k_{\rm B}T_{I}}{m} \left[-\frac{\omega_{1}}{\omega_{0}^{2}} \sin(2\omega_{1} t) 
 + \frac{\omega_{1}}{\omega_{1}^{2}} \sin(2\omega_{1} t) \right] \\
& = \omega_1 \sin(2\omega_1
t)\left[\frac{k_{\rm B}T_{I}}{m}\left(\frac{-1}{\omega_{0}^{2}}+\frac{1}{\omega_1^{2}}\right)\right] \;.
\end{align}
Here, we exclude the case that $\omega_{0} = \omega_{1}$, because this suggests no changes in the 
potential by the femtosecond-laser pulse and also the case $\omega_{1} = 0$, which would be a 
special case in which the potential would vanish after the excitation causing Supplementary Eq.\ \ref{eq:u_square} 
to be no longer a valid solution in that case. Otherwise, the expression is zero for 
time $t_{\rm max}=\pi/2\omega_1 = 1/4\nu_1$, which is the same expression we found for 
$t_2$ in the main text for the full MD simulation. Now we found that at $t_{\rm max}=\pi/{2\omega_1} 
= 1/4\nu_1$ the mean-square displacement $\langle u^{2}(t)\rangle $ reaches its maximum. 
In physical terms, this marks the turning point of the laser-induced atomic motion. Following 
our proposed excitation scheme we are going to apply now a second pulse. The displacement 
and velocity at $t_{\rm max}=\pi/2\omega_1$ are 
\begin{align}
u(t_{\rm max}) &= u_{0} \underbrace{\cos\left(\omega_1\frac{\pi}{2\omega_1}\right)}_{=0}+\frac{v_{0}}{\omega_1} \underbrace{\sin\left(\omega_1\frac{\pi}{2\omega_1}\right)}_{=1} \notag \\
&= \frac{v_{0}}{\omega_1}  \\
v(t_{\rm max}) &= -u_{0}\omega_1\sin \left(\omega_1 \frac{\pi}{2\omega_1}\right)+v_{0}\cos\left(\omega_1
\frac{\pi}{2\omega_1}\right) \notag \\
&= -u_{0}\omega_1
\end{align}

These are now initial values for the second pulse:  
\begin{align*}
v_{2} &= v(t_{\rm max}) =-u_{0}\omega_1 \\
u_{2} &= u(t_{\rm max}) =\frac{v_{0}}{\omega_1} \; .
\end{align*}
For simplicity we substitute $t' = t-t_{\rm max}$ for times larger than $t_{\rm max}$. The phonon
frequency after the second laser pulse changes to $\omega_2$. In analogy to Supplementary Eq.\ \ref{eq:u_square} 
we get:
\begin{align}
u_{2}(t') &= u_{2}\cos(\omega_{2}t')+\frac{v_{2}}{\omega_{2}}\sin(\omega_{2}t') \notag \\
&= \frac{v_{0}}{\omega_1}\cos(\omega_{2}t')-u_{0}\frac{\omega_1}{\omega_{2}}\sin(\omega_{2}t') \\
v_{2}(t') &= -u_{2}\omega_{2}\sin(\omega_{2}t')+\frac{v_{2}}{\omega_{2}}\omega_{2}\cos(\omega_{2}t') \notag \\
&=
-\frac{\omega_{2}}{\omega_1}v_{0}\sin(\omega_{2}t')-u_{0}\omega_1\cos(\omega_{2}t') \; .
\end{align} 
The equation of $\langle u_{2}^{2}(t') \rangle$ is equivalent to Supplementary Eq.\ \ref{eq:msd_harmonic}. 
Following the same mathematical steps we obtain
\begin{equation}
\langle  u_{2}^{2}(t') \rangle 
= \cos^{2}(\omega_{2}t') \frac{1}{\omega_1^{2}}\frac{k_{\rm B}T_{I}}{m}+\sin^{2}(\omega_{2}t')\frac{\omega_1^{2}}{\omega_{2}^{2}}\frac{k_{\rm B}T_{I}}{m\omega_{0}^{2}} \;.
\end{equation}
The change of the phonon frequency $\omega_1$ after the first laser pulse can be expressed by 
$\omega_1 = \alpha \omega_0$ with $\alpha$ a scalar factor. Accordingly, the change in
the phonon frequency after the second pulse can be described by $\omega_{2} = \beta \omega_0$. 
This leads to 
\begin{equation}
\label{msd_2nd_pulse}
\langle u_{2}^{2}(t') \rangle
= \cos^{2}(\omega_{2}t') \frac{1}{\alpha^{2}\omega_{0}^{2}}\frac{k_{\rm B}T_{I}}{m}+\sin^{2}(\omega_{2}t')\frac{\alpha^{2}\omega_{0}^{2}}{\beta^{2}\omega_{0}^{2}}\frac{k_{\rm B}T_{I}}{m\omega_{0}^{2}} \;.
\end{equation}
Now we would like to analyse the behavior after the second pulse. In order to obtain the 
maximal mean-square displacement $\langle u_{2}^{2}(t') \rangle$ we again try to find conditions for 
which     
\begin{equation}
\frac{\partial\langle u_{1}^{2}(t')\rangle }{\partial t'} =  0 \; . 
\end{equation}
Assuming $\omega_{2}$ is not zero, we find
\begin{align}
0 = \frac{k_{\rm B}T_{I}}{m}\sin(2\omega_{2}t')\left[\frac{\omega_1^{2}}{\omega_{2}^{2}\omega_{0}^{2}}-\frac{1}{\omega_1^{2}}\right] \; .
\end{align}
The expression on the right is zero if one of the factors is zero. 
Therefore, in analogy to the situation after the first pulse we get 
\begin{align}
\sin(2\omega_{2}t') = 0 \\
\Rightarrow t'_{\rm max} = \frac{\pi}{2\omega_2} \; .
\end{align}
More interesting is the case for when the second factor in brackets is zero. 
We get:
\begin{align}
0&=\left[\frac{\omega_1^{2}}{\omega_{2}^{2}\omega_{0}^{2}}-\frac{1}{\omega_1^{2}}\right] = \left[\frac{(\alpha \omega_0)^{2}}{(\beta\omega_{0})^{2}\omega_{0}^{2}}-\frac{1}{
(\alpha\omega_0)^{2}}\right] \\
&= \left[\frac{\alpha^{2}}{\beta^2\omega_{0}^{2}}-\frac{1}{
\alpha^2\omega_0^{2}}\right]\\
\frac{\alpha^{2}}{\beta^2\omega_{0}^{2}} &= \frac{1}{\alpha^2\omega_0^{2}}  \Rightarrow \alpha^{4} = \beta^2
\end{align}

Now, we get a relation for the phonon softening parameters $\alpha$ and $\beta$ 
\begin{equation}
\label{eq:al_be}
\alpha =  \sqrt{\beta} \; . 
\end{equation}
If now inserted into Supplementary Eq.\ \ref{msd_2nd_pulse} we obtain 
\begin{equation}
\label{msd_2nd_max}
\langle u_{2}^{2}(t')\rangle 
= \frac{1}{\alpha^{2}\omega_{0}^{2}}\cdot\frac{k_{\rm B}T_{I}}{m} \;,
\end{equation}
which is surprisingly no longer time-dependent. Moreover, it is equivalent to the 
mean-square displacement of an equilibrated system in the new, laser-changed potential.
This means after the second pulse the atomic motion is frozen, since it came at a time when the mean-square displacement already had the appropriate equilibrium value for $\omega_2$. For the case of an 
ensemble of uncoupled phonon modes in the harmonic potential we showed that our 
excitation scheme can even stop phononic motions. We note that this is an extract 
of the analysis made in Ref.\ \cite{Zier_PhD}. 

Furthermore, we computed $\alpha$ and $\beta$ for the same low-$q$ phonon for which we 
computed the potential energy in the main text. The ground state frequency  is
$\omega_0 = 2.24$ THz. After the first excitation we get $\omega_1 = 0.99$ THz. After the second excitation we have $\omega_2 = 0.75$ THz. Given these frequencies 
we obtain $\alpha = 0.44$ and $\beta = 0.33$. In the harmonic approximation we obtained 
Supplementary Eq. \ref{eq:al_be}, which would give $\alpha =  \sqrt{\beta} = 0.57$. This differs somewhat 
from our findings because of anharmonicities, but also indicates that the bond-softening 
resulting from our laser-pulse scheme does not reach these optimal values for complete 
pausing. Yet-longer-lived pausing of phononic motions may be achieved with other pulses 
or other materials that approach $\alpha =  \sqrt{\beta}$ more closely.

In addition, we used the harmonic approximation framework to show that our timed-pulse 
excitation scheme is in fact extracting energy from the system. In more detail, we compared 
the energy after a single pulse that induces a phonon frequency change from $\omega_0$ to 
$\omega_2$ with the energy after two timed pulses inducing the same change. Following 
similar steps as described above for the displacements, we obtain for the velocities 
after a single pulse to $\omega_2$:
\begin{equation}
\langle  v_{2}^{2}(t, \omega_2) \rangle 
= \frac{k_{\rm B}T_{I}}{m} \left[ \frac{\omega_2^{2}}{\omega_0^{2}}\sin^{2}(\omega_2 t) 
+ \cos^{2}(\omega_2t) \right] \;.
\label{eq:velo_om2}
\end{equation}
For the velocities after our excitation scheme with the two timed pulses we get:
\begin{equation}
\langle  v_{2}^{2}(t, \omega_1 \rightarrow \omega_2) \rangle 
= \frac{k_{\rm B}T_{I}}{m} \left[ \frac{\omega_2^{2}}{\omega_1^{2}}\sin^{2}(\omega_2 t) 
+ \frac{\omega_1^{2}}{\omega_0^{2}} \cos^{2}(\omega_2t) \right] \;.
\label{eq:velo_om1_om2}
\end{equation}
We can now compute the corresponding energies. In general, the energy $E$ of a harmonic 
oscillator is given by:
\begin{equation}
\label{eq:energy}
E = \frac{1}{2} m \omega^2 \langle u^{2}(t)\rangle  + \frac{1}{2} m \langle v^{2}
(t)\rangle \;.
\end{equation}
It becomes already clear here that an instantaneous frequency change alters the energy, more precisely 
the potential energy. For bond-softening materials like silicon, energy is extracted from the 
ionic system, whereas for bond-hardening materials, e.g., gold \cite{Descamps2024} and magnesium 
\cite{Grigoryan}, energy is pumped into the system. Using Eqs.\ \ref{eq:reult_msd_one_pulse} 
and \ref{eq:velo_om2} in Supplementary Eq.\ \ref{eq:energy} we obtain the energy $E_{\omega_2}$ for a single 
pulse excitation that induces a phonon frequency change from $\omega_0$ to $\omega_2$: 
\begin{equation}
E_{\omega_2} = \frac{1}{2} m \omega_{2}^{2} \left( \cos^{2}(\omega_2
t) \frac{k_{\rm B}T_{I}}{m\omega_{0}^{2}}+\sin^{2}(\omega_2
t)\frac{k_{\rm B}T_{I}}{m\omega_2^{2}} \right) + \frac{1}{2} m \frac{k_{\rm B}T_{I}}{m} \left( 
\frac{\omega_2^{2}}{\omega_0^{2}}\sin^{2}(\omega_2 t) + \cos^{2}(\omega_2t) \right)
\label{eq:Energy_om2} \:, 
\end{equation}
which can be reduced to
\begin{equation}
E_{\omega_2} = \frac{1}{2} k_{\rm B}T_{I} \left(\frac{\omega_2^{2}}{\omega_0^{2}} + 1 \right)
\label{eq:res_Energy_om2} \:. 
\end{equation}
By using the relation of the softening parameters, Supplementary Eq.\ \ref{eq:al_be}, we get:
\begin{equation}
E_{\omega_2} = \frac{1}{2} k_{\rm B}T_{I} \left( \alpha^4 + 1 \right)
\label{eq:fin_Energy_om2} \:. 
\end{equation}
Similarly, we computed the energy $E_{\omega_1 \rightarrow \omega_2}$ after the double-pulse:
\begin{equation}
E_{\omega_1 \rightarrow \omega_2} = \frac{1}{2} k_{\rm B}T_{I} \left(\frac{\omega_2^{2}}{\omega_1^{2}} + \frac{\omega_1^{2}}{\omega_0^{2}} \right)
\label{eq:res_Energy_om1_om2} \:. 
\end{equation}
Using again the relation in Supplementary Eq.\ \ref{eq:al_be}, this equation reduces to:
\begin{equation}
E_{\omega_1 \rightarrow \omega_2} = \frac{1}{2} k_{\rm B}T_{I} \left(2\alpha^2 \right)
\label{eq:fin_Energy_om1_om2} \:. 
\end{equation}

In a next step we compute the energy difference:
\begin{align}
\Delta E &= E_{\omega_2} - E_{\omega_1 \rightarrow \omega_2} \\
&= \frac{1}{2} k_{\rm B}T_{I} \left( \alpha^4 + 1 \right) - \frac{1}{2} k_{\rm B}T_{I} \left(2\alpha^2 \right)  \\
&= \frac{1}{2} k_{\rm B}T_{I} \left[ \left( \alpha^4 + 1 \right) - \left(2\alpha^2 \right) \right] \\
&= \frac{1}{2} k_{\rm B}T_{I} \left(  \alpha^2 - 1 \right)^2 \:.
\end{align}
For $\alpha = 1$ no changes are induced in the system and the energy difference is zero. 
As a consequence of phonon softening, we observe $0 < \alpha < 1$, which causes the 
energy following our excitation scheme to be always smaller than the energy of a single 
pulse excitation to the same excitation level. We note that our derivations are 
in principle also valid for materials that show phonon-hardening $\alpha > 1$ after an 
femtosecond-laser excitation.

\section*{\textbf{Supplementary Note 2:}\label{sec:BraggPeak}Additional Bragg Peak intensities}

The normalized $(111)$ Bragg peak intensities shown in the main text are computed 
by 
\begin{equation}
I_{hkl}  (t)= \frac{1}{N^2/2} \left| \sum^{N}_{j = 1} e^{ - 2 \pi i \;\left[hx_{j}(t) + kx_{j}(t) + lx_{j}(t)\right]} \right|^2 \; ,
\label{eq:1}
\end{equation}
with $(hkl)= (111)$, $N=1000$ the number of silicon atoms in the used supercell, 
and $x_{j}(t),y_{j}(t),z_{j}(t)$ the time-dependent position of the $j$-th atom in units 
of the lattice parameter. The results shown here are averaged over the trajectories of all 
ten independent runs. Suppl. Figs.\ \ref{fig:220} and \ref{fig:311} present additional 
results for Bragg intensities other than (111), namely for $(220)$ and $(311)$, 
respectively. In all of them the effect of preserved crystal symmetry is present.

\begin{figure*}[ht!]
\includegraphics[width=0.7\textwidth]{SF_220.ps}
\caption{\label{fig:220}Time-evolution of the (220) Bragg peak intensities for 
the same laser-induced electronic entropies as discussed in the main text.}
\end{figure*}

\begin{figure*}[ht!]
\includegraphics[width=0.7\textwidth]{SF_311.ps}
\caption{\label{fig:311}(311) Bragg peak intensity as a function of time after 
different excitation strengths and mechanisms.}
\end{figure*}

\newpage
\section*{\textbf{Supplementary Methods:}\label{sec:EPC} Effects of optical phonon heating by electron-phonon coupling}

\begin{figure*}[ht!]
\includegraphics[width=0.7\textwidth]{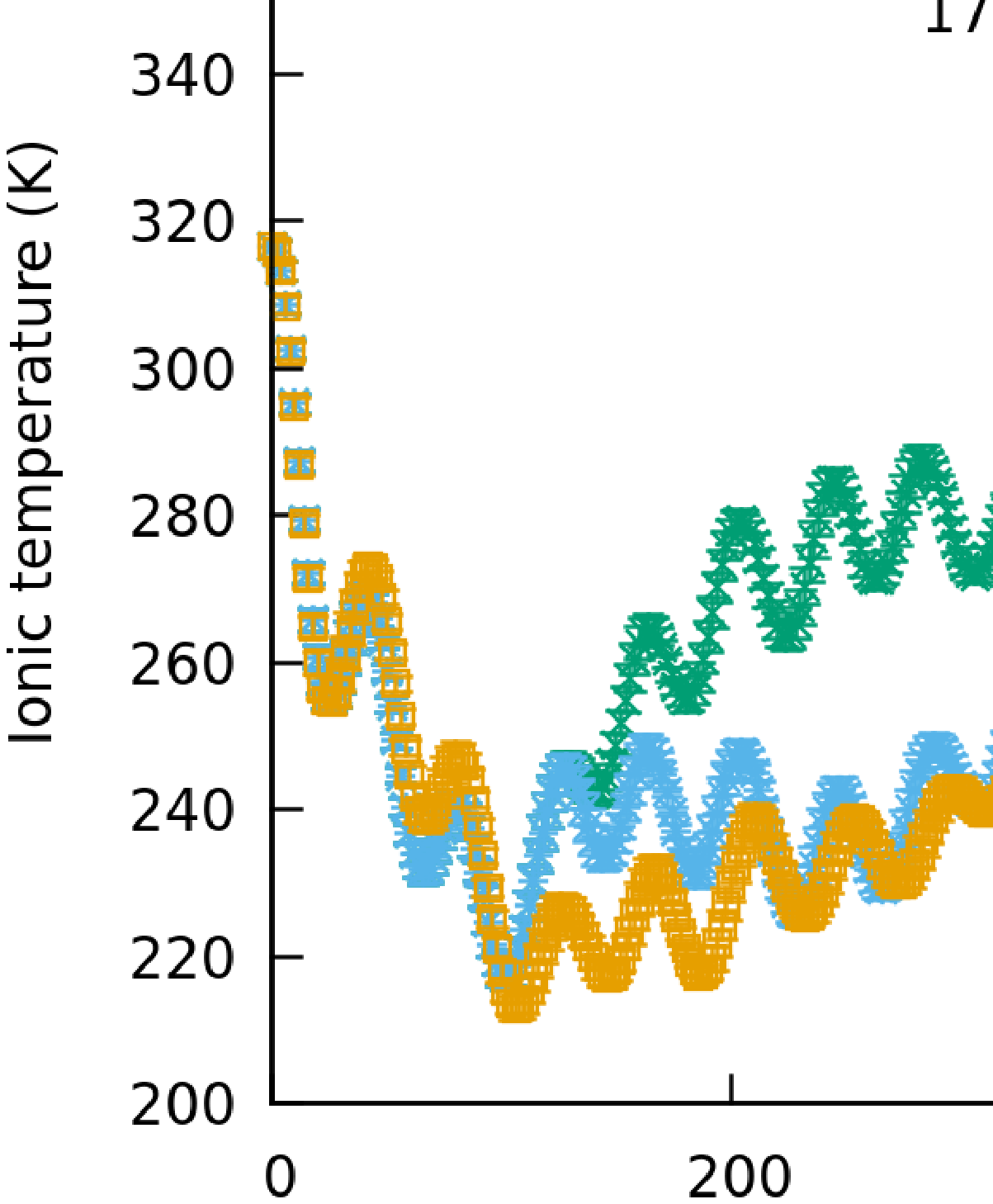}
\caption{\label{fig:epc_ion_temp} Total ionic temperature with optical phonon rescaling as a 
function of time after the excitation for the three excitations discussed in the main text.}
\end{figure*}

A crucial part within modelling laser-excited materials is the equilibration process 
between electrons and ions, the incoherent heating of phonons by the extremely hot 
electronic system. At present no all-encompassing {\it ab initio} theory is available for 
strongly laser-excited materials. Many research works have used many-body-perturbation 
theory or density-functional-perturbation-theory (DFPT), but mostly for lower-intensity 
cases or shorter times. Usually the electrons do not exceed temperatures of $3000$ K.
However, one main result of previous research is the strong and fast electron-phonon 
coupling of optical phonons, in particular low-$q$ phonons near the $\Gamma$ point 
\cite{Sadavisam,Park2020}. Their temperature can be very high, $\sim 1000$ K after 
$t = 100$ fs \cite{Sadavisam}. Through phonon-phonon coupling this energy is distributed 
within the phonon system, but mainly within the optical phonon branches. 

In this sense we implemented an effective electron-phonon coupling by heating all optical phonon 
branches (according to the selection in the main text Suppl. Fig.\ 3(a)) via velocity rescaling. 
Therefore, the total ionic velocity $\vec{v}$ is projected on the phonon eigenvectors $\vec{e}_j$ to 
obtain the phonon velocity $\vec{v}_j$:
\begin{equation}
\label{eq:v_project}
\vec{v}_{j} = v_{j}\vec{e}_j \;,
\end{equation} 
with $v_{j} = \vec{v} \cdot \vec{e}_j$. In this way the total velocity is given by:
\begin{equation}
\label{eq:v_total}
\vec{v} = \sum_{j=1}^{3N} \vec{v}_j \;,
\end{equation}
with $3N$ the number of phonons in the system. Extending the sum to the different phonons, 
we get:
\begin{equation}
\label{eq:v_decomp}
\vec{v} = \sum_{a=1}^{N_{\rm ac}} \vec{v}_a + \sum_{m=1}^{N_{\rm mi}} \vec{v}_m + \sum_{o=1}^{N_{\rm 
opt}} \vec{v}_o \;,
\end{equation}
where $N_{\rm ac}, N_{\rm mi}$, and $N_{\rm opt}$ are the numbers of acoustic, mixed, and optical 
phonons as decomposed in the main text, respectively. In order to effectively heat the optical 
phonons, we multiply the last term in each timestep of our simulation by $\epsilon$:
\begin{equation}
\label{eq:v_decompII}
\vec{v} = \sum_{a=1}^{N_{\rm ac}} \vec{v}_a + \sum_{m=1}^{N_{\rm mi}} \vec{v}_m + \epsilon \sum_{o=1}
^{N_{\rm opt}} \vec{v}_o \;.
\end{equation}
The connection of the ionic velocity to the ionic temperature of the system is given by the 
equipartition relation:
\begin{equation}
\label{eq:temp1}
\langle E_{\rm kin} \rangle = \frac{3}{2} m \langle v^2 \rangle = \frac{3}{2} \left( N - 1 \right)
k_{\rm B} T_I \;.
\end{equation}
Reference \cite{Sadavisam} reported an temperature increase from $300$ K to roughly $700$ K 
in the optical phonons within the first $2$ ps after the excitation. In case the electron-phonon 
coupling is much faster and stronger than reported, we exaggerated the scaling factor in order to get 
an upper limit. Instead of the ionic temperature, we increased the optical phonon velocity by a 
factor of $7/3$ in the first ps. Therefore, the scaling factor $\epsilon$ for each timestep is  
\begin{equation}
\label{eq:v_rescaleII}
\epsilon = \sqrt[500]{7/3} \approx  1.001696 \;,
\end{equation} 
where $500$ is the number of timesteps to reach $1$ ps simulation time. Without phonon-phonon 
interactions or other external perturbations like the laser, the optical phonon temperature of 
initially $T_I = 300$ K is rescaled by a factor of $\epsilon^2$ to roughly $T= 1715$ K within the 
first ps. Again, this is more than twice the temperature reported in \cite{Sadavisam} after 
$t = 2$ ps. Acoustic and mixed phonons are not altered but could gain energy by phonon-phonon 
interactions, which are completely included in CHIVES.

Suppl. Figure \ref{fig:epc_ion_temp} shows the total ionic temperature as a function of time for all 
three excitations mentioned in the main text. The initial dip and the oscillations are induced 
by the laser. The ensuing increase comes from the modelled heating of the ions. By analogy 
with the main text we decomposed the total ionic temperature. In Suppl. Fig.\ \ref{fig:epc_decom47} 
the time evolution of the partial ionic temperatures is shown for an excitation to $S_e = 5.05$ 
mHa K$^{-1}$ ($T_e = 14 842$ K).  
\begin{figure*}[ht!]
\includegraphics[width=0.7\textwidth]{EPC_decomp_47mHa.ps}
\caption{\label{fig:epc_decom47} Decomposed ionic temperature with optical phonon rescaling 
for $T_e = 14842$ K ($S_e = 5.05$ mHa K$^{-1}$) on the three ranges (optical, mixed, acoustic) defined 
in the main text.} 
\end{figure*}
As intended the temperature of the optical branch shows on average a monotonic increase. Besides that, 
energy is transferred significantly to the mixed phonons after around $500$ fs, after which a slight 
increase is visible. Compared to the non-rescaled case, which settled around $250$ K within our 
simulation time, the temperature here was raised to above $300$ K. There is basically no change in the 
acoustic phonon temperature compared to the non-rescaled case. For the excitation to $T_e = 17368$ K 
($S_e = 6.14$ mHa K$^{-1}$) in a single pulse  (Suppl. Fig.\ \ref{fig:epc_decomp47_55} (a)) all three decomposed 
temperatures are increasing within the simulation time. For comparison, all three temperatures stayed 
close to or below $250$ K before rescaling. Here, even the acoustic phonon's temperature is slightly 
increased by roughly $50$ K. We note that in this case the overall temperature in the optical 
phonons reaches only $400$ K, whereas for the weaker excitation to $T_e = 14 842$ K (Suppl. Fig.\ 
\ref{fig:epc_decom47}) it exceeds $450$ K. This is attributed to significantly more phonon-phonon 
interactions, also seen by the heating of the other decomposed temperatures. 
\begin{figure*}[ht!]
     \centering
     \begin{subfigure}[b]{0.48\textwidth}
         \centering
         \includegraphics[width=\textwidth]{EPC_decomp_55mHa.ps}
          \caption{}
     \end{subfigure}
     \hfill
     \begin{subfigure}[b]{0.48\textwidth}
         \centering
         \includegraphics[width=\textwidth]{EPC_decomp_47to55mHa.ps}
          \caption{}
     \end{subfigure}
\caption{\label{fig:epc_decomp47_55} The ionic temperature decomposition with optical phonon 
rescaling into the three phonon ranges defined in the main text for (a) the single-pulse 
excitation to $T_e = 17368$ K ($S_e = 6.14$ mHa K$^{-1}$) and (b) for the double-pulse 
excitation. }
\end{figure*}
\begin{figure*}[ht!]
\includegraphics[width=0.7\textwidth]{EPC_MSD.ps}
\caption{\label{fig:epc_MSD} Averaged mean-square atomic displacement as a 
function of time after the femtosecond-laser pulse for all three excitations 
discussed in the main text, under the influence of an effective electron-phonon 
coupling.}
\end{figure*}
After our double-pulse excitation scheme (Suppl. Fig.\ \ref{fig:epc_decomp47_55}(b)) we observe a similar 
increase of the optical temperature as in Suppl. Fig.\ \ref{fig:epc_decom47} and only a smaller increase 
in the mixed temperature towards the end of our simulation. The temperature of the acoustic part 
is basically not affected. Remarkably, towards the end of our simulation its temperature is 
{\it even smaller} than in the case without effective electron-phonon coupling. Since those 
phonons are of immense importance to our excitation scheme, this result indicates a robustness 
of the pausing mechanism to effective ultrafast optical phonon heating by electron-phonon 
coupling. 

This trend can also be seen in the mean-square atomic displacement (Suppl. Fig.\ \ref{fig:epc_MSD}). 
After the double-pulse the mean-square atomic displacement remains constant 
well below $0.1$ \AA$^2$, whereas the values for a single pulse excitation increase above 
$0.5$ \AA$^2$, a much larger effect than in the case without effective electron-phonon coupling. 
For completeness we also show the results for the Bragg peak intensities for the ($111$) and 
($220$) directions (Suppl. Fig.\ \ref{fig:epc_bragg}) which show that our excitation scheme prevented 
loss of $30$\% of the initial intensity.  
\begin{figure*}[ht!]
     \centering
     \begin{subfigure}[b]{0.48\textwidth}
         \centering
         \includegraphics[width=\textwidth]{EPC_SF_111.ps}
          \caption{}
     \end{subfigure}
     \hfill
     \begin{subfigure}[b]{0.49\textwidth}
         \centering
         \includegraphics[width=\textwidth]{EPC_SF_220.ps}
          \caption{}
     \end{subfigure}
\caption{\label{fig:epc_bragg} Averaged time-dependent Bragg intensity for (a) the ($111$) 
and (b) the ($220$) direction after the different excitations described in the main text, with 
an additional effective electron-phonon coupling. Vertical widths show standard deviations. }
\end{figure*} 
In summary, our excitation scheme that causes pausing of ionic motions is resilient to an 
optical phonon heating by electron-phonon coupling even at a level that is twice as fast and more 
than twice as strong as reported in previous works \cite{Sadavisam}.

\section*{\textbf{Supplementary Note 3:}\label{sec:PCF}Pair-correlation function}

\begin{figure*}[t]
\includegraphics[width=0.5\textwidth]{pair_corr_fct.ps}
\caption{\label{fig:pcf} Averaged pair-correlation function $g(r)$ of the ground state 
initialized with $T =315$ K compared to the system at $t=1$ ps after: (top) a single pulse 
excitation to $T_e = 14 842$ K ($S_e = 5.05$ mHa K$^{-1}$), (middle) our proposed double-pulse 
excitation to a final $S_e = 6.14$ mHa K$^{-1}$, and (bottom) a single-pulse excitation to 
$T_e=17 368$ K ($S_e = 6.14$ mHa K$^{-1}$).}
\end{figure*}

In order to characterize specific changes in the crystalline structure after applying 
our excitation mechanism, we computed the pair-correlation function $g(r)$ 
directly from the time-dependent atomic positions, by:
\begin{equation}
    g (r) = \sum_{i,j} \frac{G(r-r_{ij})}{2 \pi r^2 N} \; .
  \label{eq:direct_pcf}
\end{equation}
In this equation, we sum over Gaussians $G$ with a FWHM of 0.1 Å. $r_{ij}$ 
describes the distance between atom $i$ and $j$. Suppl. Figure \ref{fig:pcf} shows $g(r)$ 
averaged over $10$ independent runs at $t=1$ ps for all induced electronic 
temperatures studied in this work in comparison to the ground state at $315$ K. 
The single-pulse excitation below the ultrafast melting threshold 
($T_e = 14 842$ K, $S_e = 5.05$ mHa K$^{-1}$, Suppl. Fig.\ \ref{fig:pcf}, top panel, 
green solid line) causes relatively small changes, mainly in the peak height, but 
not in the structure of the function. For the single excitation to $T_e=17 368$ K, 
$S_e=6.14$ mHa K$^{-1}$ (Suppl. Fig.\ \ref{fig:pcf}, bottom panel, orange solid line) with a 
single pulse completely different characteristics are obtained after $1$ ps. The 
first, main, peak shifted and the structure at longer interatomic distances is 
completely washed out to an almost constant value around $1$, a clear sign for 
a crystal symmetry loss. After $1$ ps in the controlled phase, however, several 
characteristics are preserved in the presence of extremely hot electrons. In 
particular, the first peaks remained close to the ground state. Since the first 
peak corresponds to the nearest-neighbor distance, we can conclude that the 
short-range symmetry is preserved in this case. However, for longer interatomic 
distances characteristics are slightly washed out, implying the loss of long-range order.

\newpage
\section*{\textbf{Supplementary Note 4:}\label{sec:SF} Structure function}

An additional accessible quantity by experiments is the structure function, which 
we computed in accordance with \cite{Zier17}. We used \cite{Lin2006}:
\begin{equation}
I(q,t) = \frac{\sum_\mathbf{q'} I_\mathbf{q'}(t) \;G\left( q - q'\right)}{\sum_\mathbf{q'} G\left( q - q'\right)} \; ,
\label{eq:structurefunction}
\end{equation}
where  $I_{hkl}(t)$ is described by Supplementary Eq.\ (\ref{eq:1}) and $G$ is again a
Gaussian, with FWHM of $0.2 \;\text{\AA}^{-1}$, which we used to smooth the 
peak intensities. We computed this quantity for each of the individual runs 
and averaged over $N_{\rm run}$. Suppl. Fig.\ \ref{fig:SF} shows the results for the controlled phase 
at $1$ ps after the double pulse excitation and for the ground state. Both data
sets are averaged over our ten independent MD-simulation runs. Although plotted 
with plus-minus the standard deviation, their error is too small to be visible 
in this figure. In order to make the difference more visible, we additionally 
plotted the difference of both curves. Whereas all main peaks with corresponding 
($hkl$) decrease in the controlled phase, all intensities between main peaks 
increase, indicating an increased atomic displacement in the controlled phase compared to 
the ground state. 

\begin{figure*}[ht!]
\includegraphics[width=0.7\textwidth]{Struct_fct_diff.ps}
\caption{\label{fig:SF} Averaged structure function as a function of the scattering vector 
for the ground state (grey solid line) and after $1$ ps in the controlled phase 
(light blue solid line). For easier interpretation, we plotted additionally the
difference between both data sets (red solid line). }
\end{figure*}


\newpage
\section*{\textbf{Supplementary Note 5:}\label{sec:PES} Effective potential-energy surface}

In order to get insight into the microscopic mechanisms causing the different 
outputs described in the main text, we used the ability of the atoms to probe 
the underlying effective interatomic potential given by DFT. Our MD-simulations 
provide the time-dependence of the total force $\vec{F}(t)$ and the atomic 
positions $\vec{r}(t)$. By projecting those quantities on the eigenvector $\vec{e}_j$ 
of the low-$q$ phonons, we obtained for every timestep the force constant $F$ and 
the displacement $u$ in this particular phonon direction. This data $F(u)$ set is 
then fitted to a third-order polynomial of the form:
\begin{equation}
    F(u) = a u^3 + b u^2 + c u + d \; ,
\end{equation}
with real coefficients $a,b,c$ and $d$. We would note that for the 
double-pulse excitation only the time after the second pulse was fitted. The 
coefficients obtained are summarized in Table \ref{tab_coeff}.
\begin{table}[ht!]
\renewcommand{\arraystretch}{2}
\begin{tabular}{|l||r|r|r|r|}
\hline
& \multicolumn{1}{c}{$a$} \vline & \multicolumn{1}{c} {$b$} \vline & \multicolumn{1}{c} {$c$} \vline & \multicolumn{1}{c} {$d$} \vline \\
\hline
     &  \multicolumn{1}{c}{Ha/(bohr)$^4$} \vline & \multicolumn{1}{c}{Ha/(bohr)$^3$} \vline &  \multicolumn{1}{c}{Ha/(bohr)$^2$} \vline & \multicolumn{1}{c}{ Ha/(bohr)} \vline \\
\hline \hline

 Equilibrium structure 315 K  &  &  & $(-5.764      \pm 0.006)\times 10^{-3}$ &  $(-6.7 \pm 2.5) \times 10^{-6}$\\
\hline
5.05 mHa K$^{-1}$     & $(6.3 \pm 4.0)\times 10^{-6}$ & $(-3.1 \pm 0.4)\times 10^{-6}$ & $(-9.79 \pm 0.07) \times10^{-4}$ & $(-5.4 \pm 0.3)\times 10^{-7}$ \\
\hline
6.14 mHa K$^{-1}$ (double pulse)   & $(2.2 \pm 0.4)\times 10^{-6}$  & $(-2.2 \pm 0.6)\times 10^{-6}$ & $(-4.82 \pm 0.08)\times 10^{-4}$ & $(8.2 \pm 0.5)\times 10^{-5}$ \\
\hline
6.14 mHa K$^{-1}$     & $(-6.9 \pm 0.8)\times 10^{-6}$  & $(4.7 \pm 3.0)\times 10^{-6}$ & $(-2.9 \pm 0.1)\times 10^{-4}$ & $(1.4 \pm 0.9)\times 10^{-5}$  \\
\hline
\end{tabular}
\caption{\label{tab_coeff} Coefficients $a$, $b$, $c$, $d$ of the fitted 
polynomial of the interatomic forces for the equilibrium structure at 
$T=315$ K and the three electronic excitations discussed in the main text.}
\end{table}

\begin{figure*}[ht!]
\includegraphics[width=0.8\textwidth]{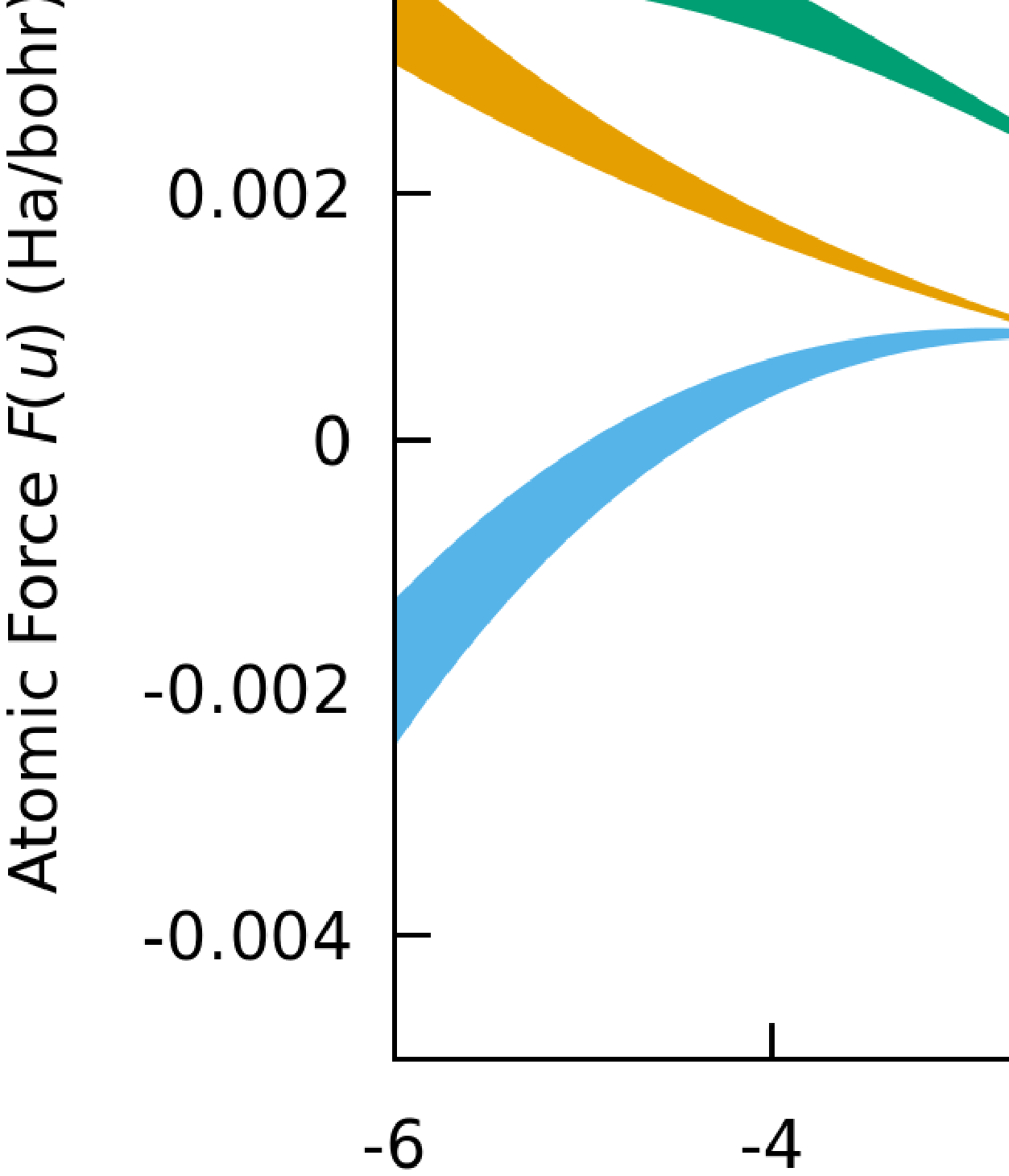}
\caption{\label{fig:error} The mean value of the polynomial fit of the 
interatomic forces for each excitation plus/minus its error by correlated 
error propagation  \cite{error_propagation}.}
\end{figure*}
 
\begin{figure*}[ht!]
     \centering
     \begin{subfigure}[b]{0.495\textwidth}
         \centering
         \includegraphics[width=\textwidth]{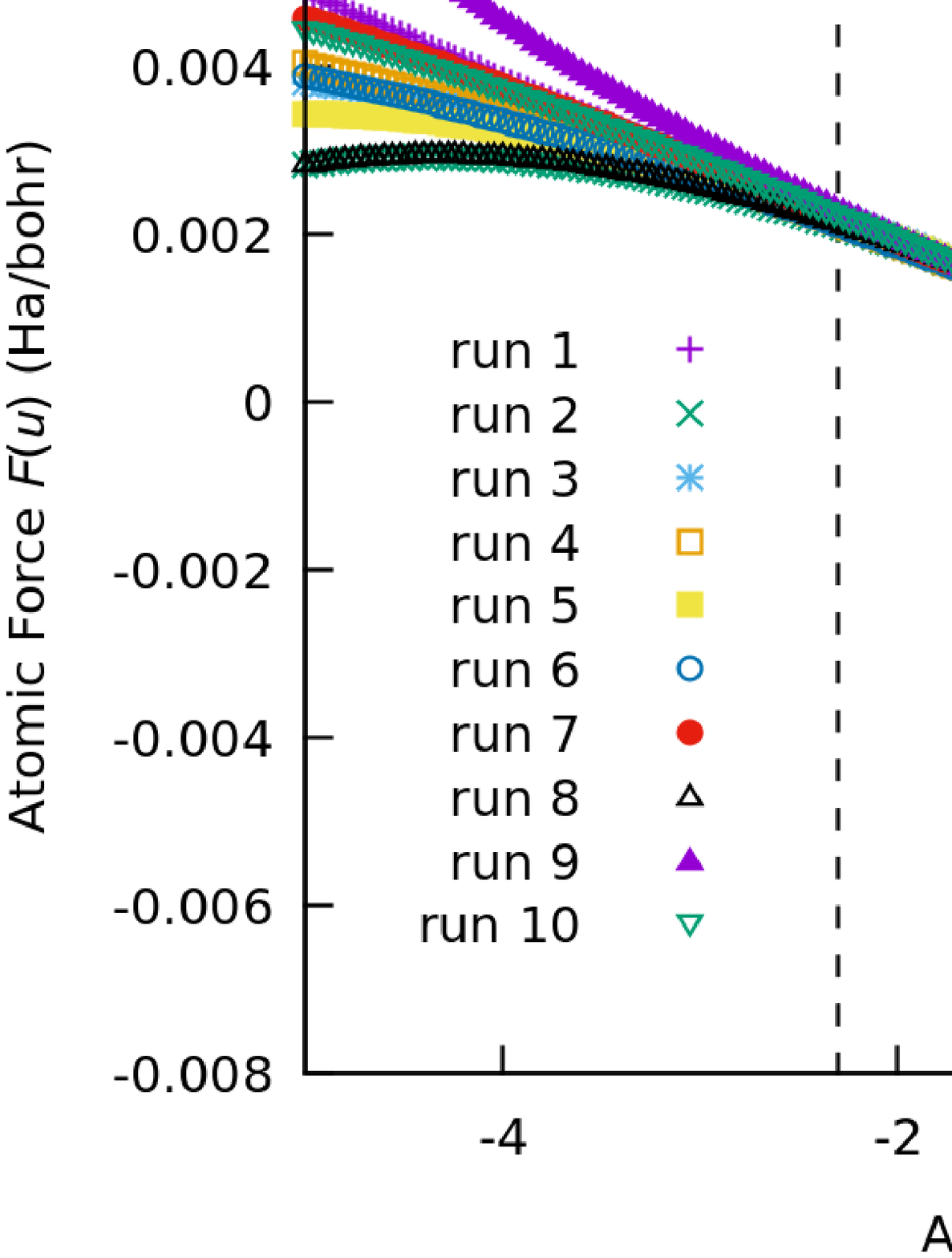}
     \end{subfigure}
     \hfill
     \begin{subfigure}[b]{0.495\textwidth}
         \centering
         \includegraphics[width=\textwidth]{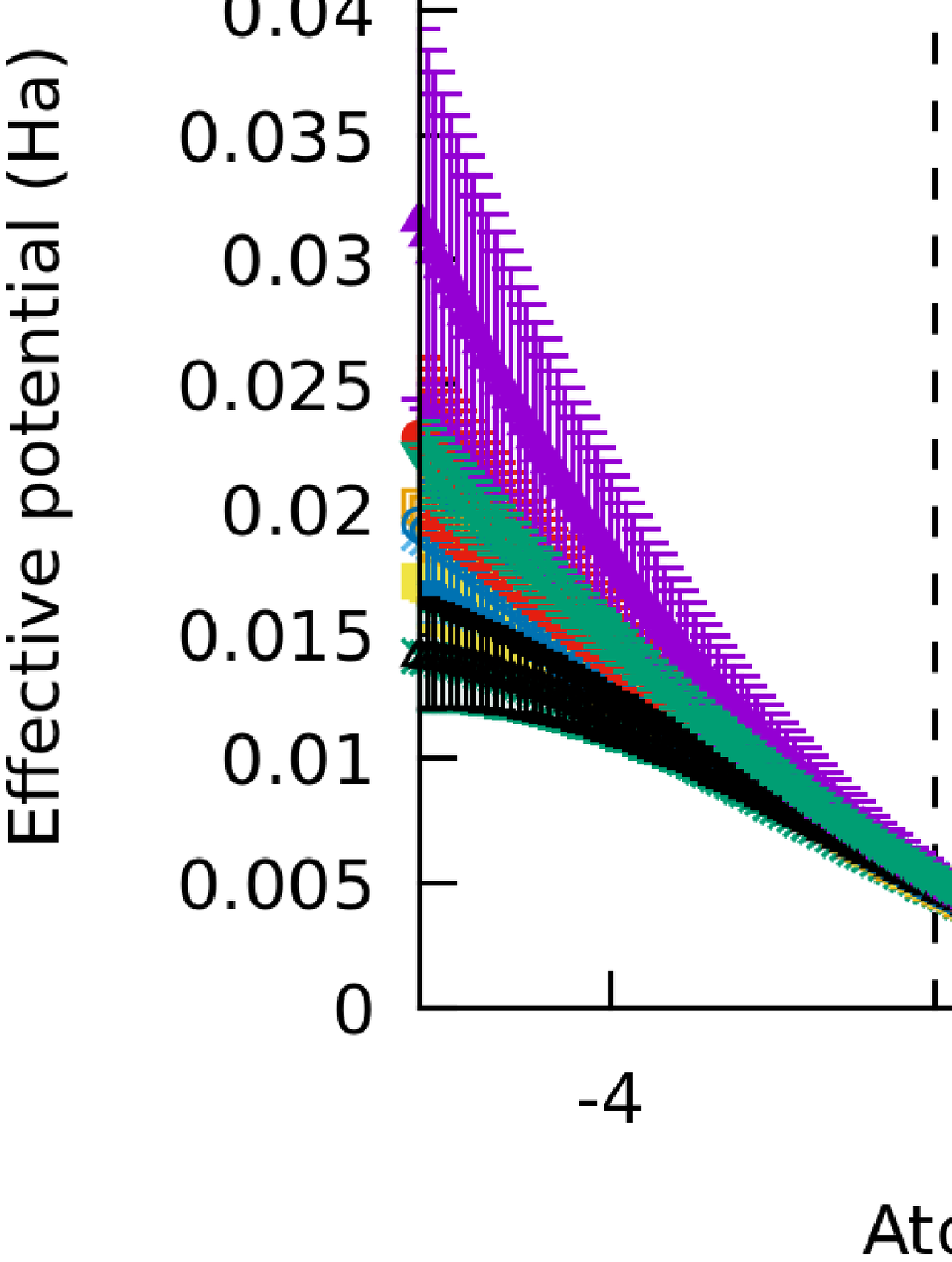}
     \end{subfigure}       
\caption{\label{fig:F_POT_47mHa} (left) The interatomic fitted forces $F(u)$ for each 
simulation run (1-10) for $S_e = 5.05$ mHa K$^{-1}$. (right) The effective potential energy for the same excitation strength. Vertical dashed lines indicate the average atomic displacement range.}
\end{figure*}

\begin{figure*}[ht]
     \centering
     \begin{subfigure}[b]{0.49\textwidth}
         \centering
         \includegraphics[width=\textwidth]{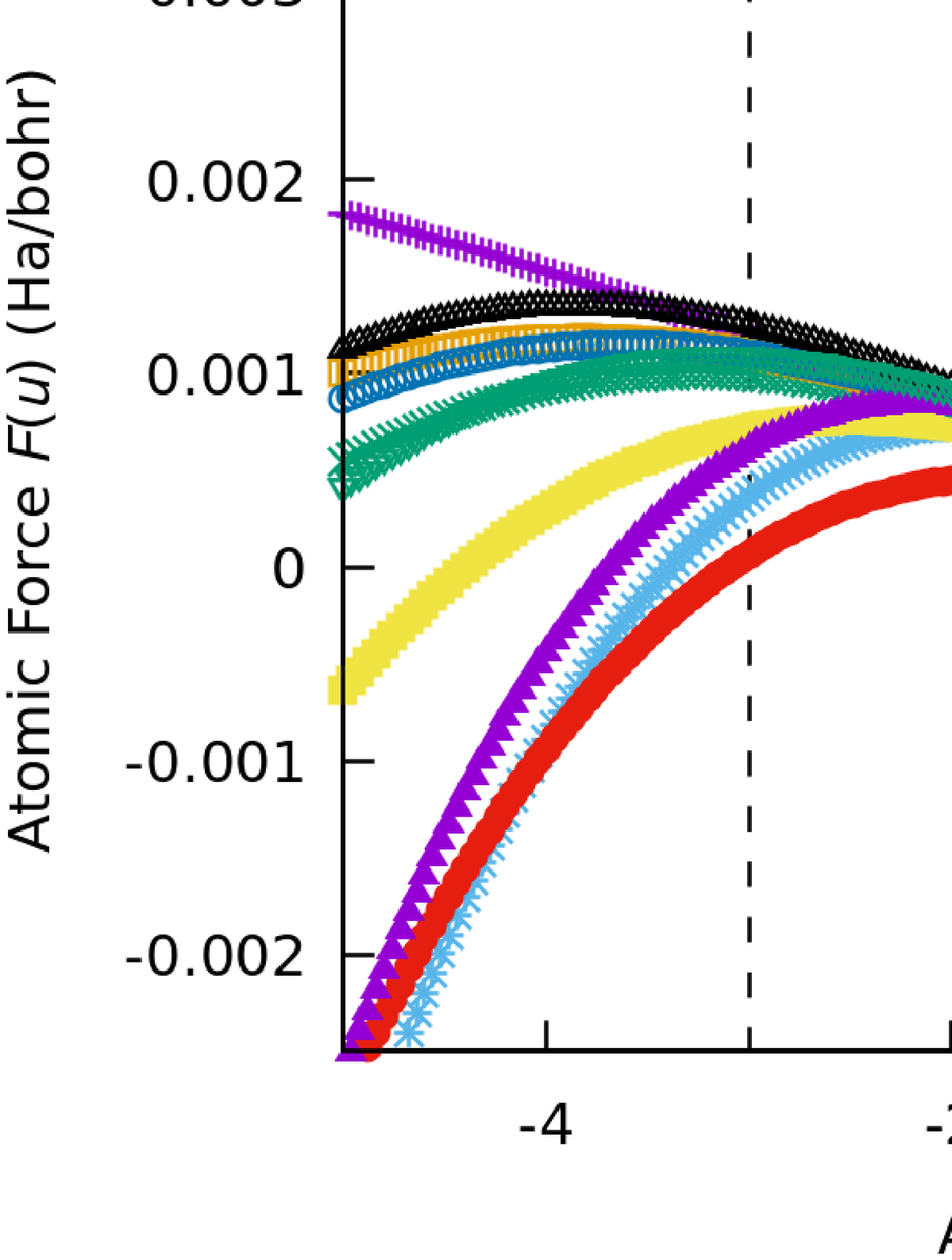}
          \caption{}
     \end{subfigure}
     \hfill
     \begin{subfigure}[b]{0.48\textwidth}
         \centering
         \includegraphics[width=\textwidth]{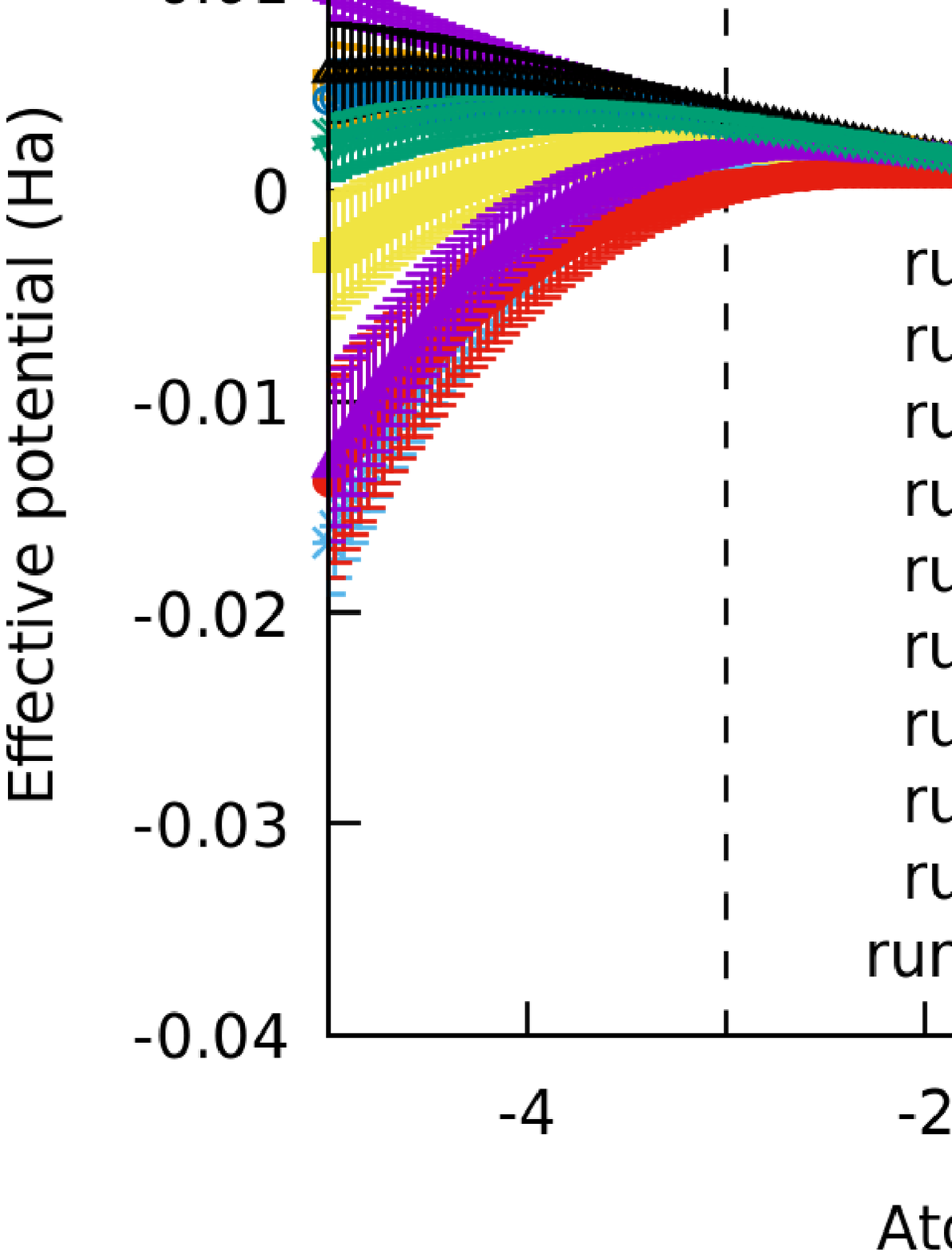}
          \caption{}
     \end{subfigure}

     \centering
     \begin{subfigure}[b]{0.49\textwidth}
         \centering
         \includegraphics[width=\textwidth]{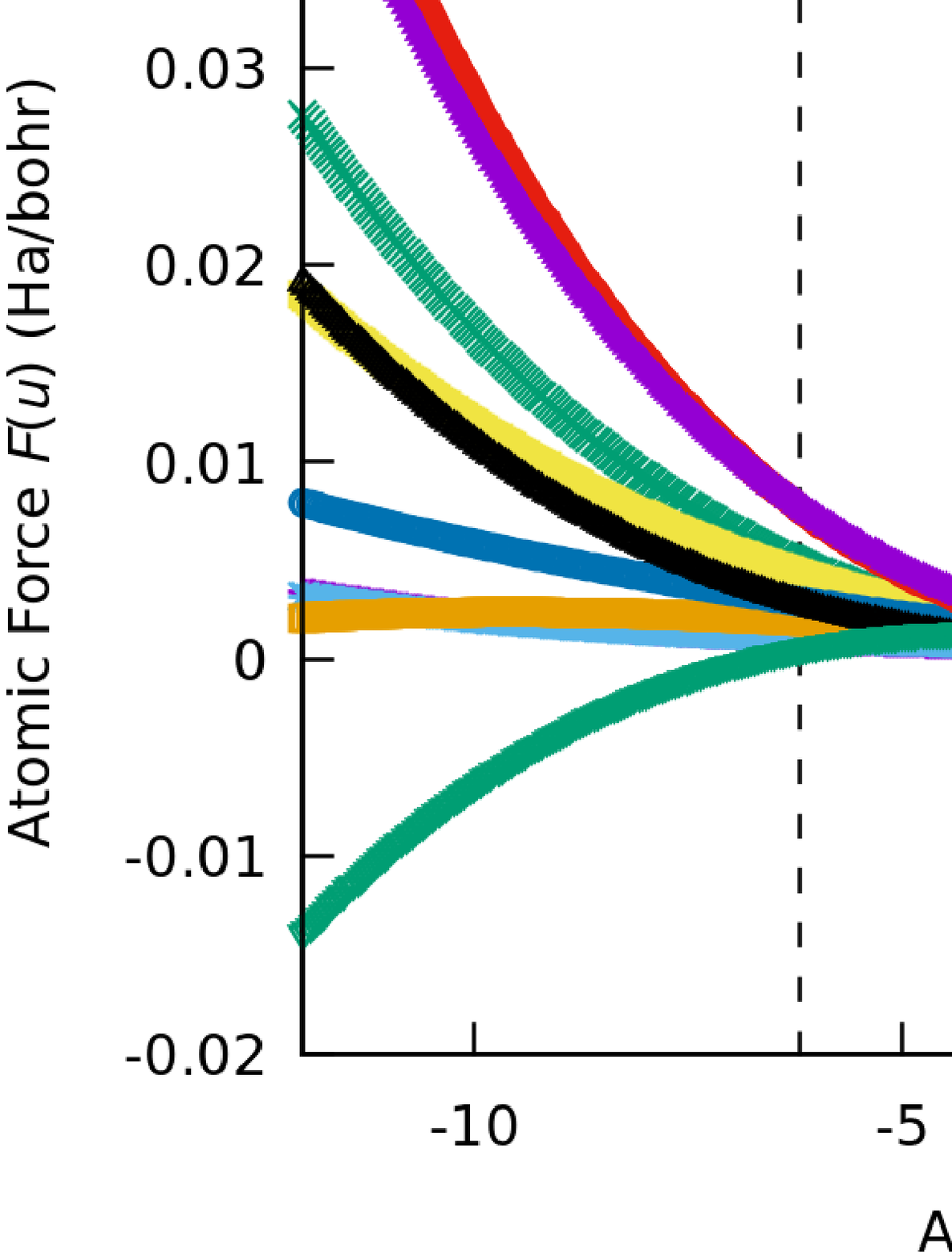}
          \caption{}
     \end{subfigure}
     \hfill
     \begin{subfigure}[b]{0.47\textwidth}
         \centering
         \includegraphics[width=\textwidth]{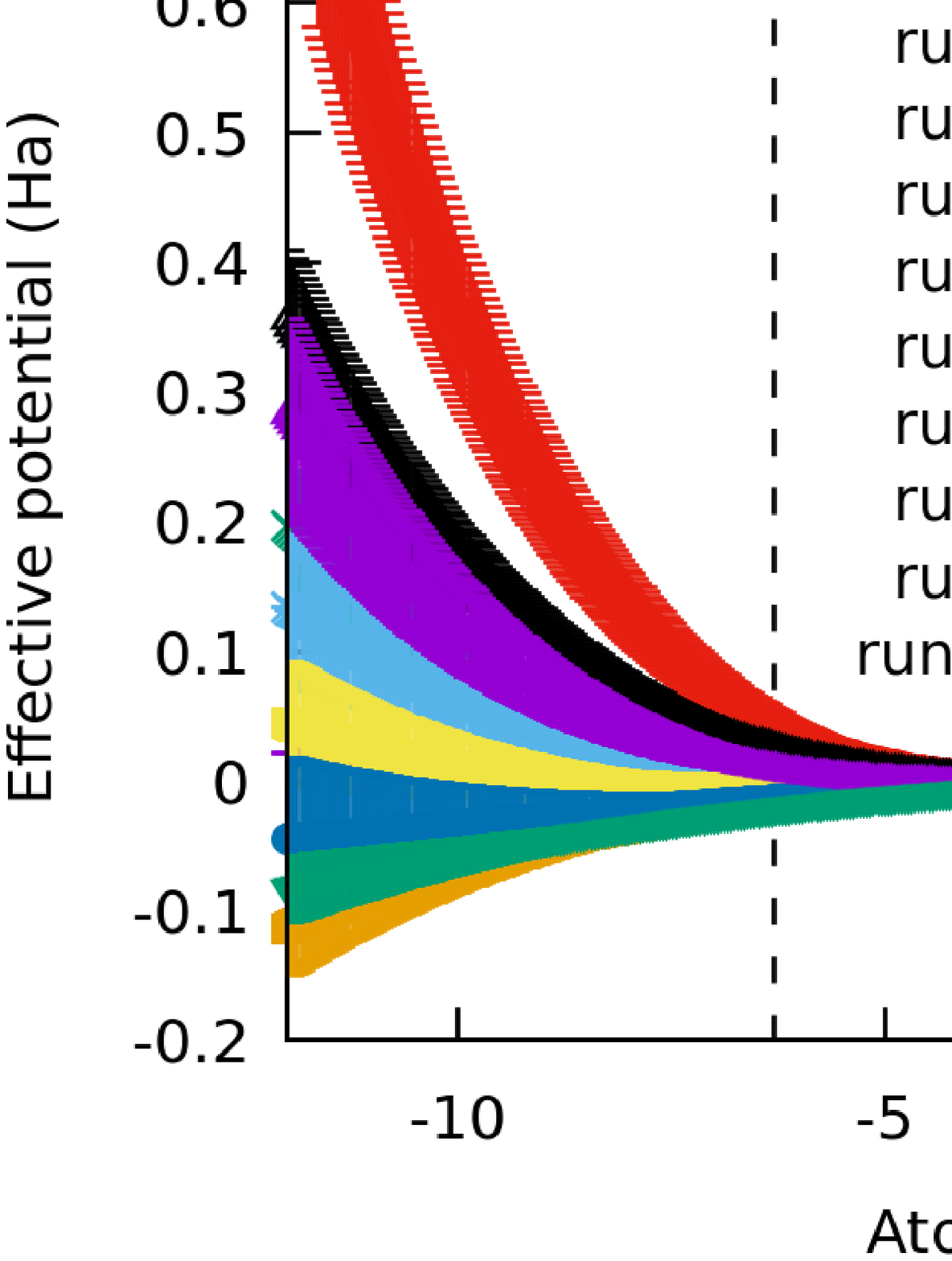}
          \caption{}
     \end{subfigure}
\caption{\label{fig:F_POT_55mHa} (a) The interatomic fitted forces $F(u)$ for the 
double-pulse excitation to $S_e = 6.14$ mHa K$^{-1}$ for each simulation run (1-10). (b) The 
effective potential energy for the same excitation strength. (c) $F(u)$ for each 
simulation run (1-10) for a single-pulse excitation to $S_e = 6.14$ mHa K$^{-1}$ for a wider 
atomic displacement range and (d) the corresponding effective potential. Vertical 
dashed lines indicate the average atomic displacement range.}
\end{figure*}
For the equilibrium structure at $315$ K a linear fit was sufficient enough to describe 
the potential. In this case the atoms do not move far enough to probe any 
anharmonicities of the potential. The polynomial fits to the interatomic forces and 
their error can be seen in Suppl. Fig.\ \ref{fig:error}.   
The displacement of atoms in the above-mentioned phonon direction varies between 
the runs, because of the different initial conditions. For instance, the direction 
$+u$ is predominant in one run and $-u$ in another one. As a consequence the 
effective atomic potential is only probed in this range of $u$, which could lead 
to different shapes of the fitted force dependencies. Therefore, we plotted the 
mean value $F(u)$ for every single run in Suppl. Figs. \ref{fig:F_POT_47mHa} -- 
\ref{fig:F_POT_55mHa} (left column) as well as the integrated effective potential 
energy (right column) for the three excitations described in the main text. With dashed 
vertical lines we indicate the average atomic displacement range.
Two different behaviors can be observed from this data. First, in the vicinity of 
$u=0$ the characteristics do not vary between the runs; the slope as well as the 
absolute value of $F$ are the same within each run. This region (ranging from $2$ 
to $5$ bohr depending on the excitation) seems to be probed by the atoms very 
accurately. Secondly, for larger $u$, we observe large differences between the 
fitted forces. This is a direct consequence of the above-mentioned fact that the 
atoms do not probe the same phase space in each run. In addition, the surrounding 
configuration is different in each run, which affects the probed effective 
interatomic force, especially for the laser-induced melting case. Therefore, 
the value for very large $u$ is not necessarily meaningful for extrapolation.
However, the slopes of $F(u)$ observed here verify that after the double-pulse 
excitation an attractive potential is still present. We note that the linear slope 
around $u=0$, corresponding to a harmonic potential, decreased 
by $91.6$\% compared to the equilibrium structure at $T=315$ K. After the single-pulse 
excitation to $S_e = 6.14$ mHa K$^{-1}$, the same slope decreased by $95.0$\% and $39.9$\% compared to the 
equilibrium state and the double-pulse excitation, respectively. Therefore, in this case 
the vicinity of $u=0$ can still be described by an attractive harmonic potential, 
which is more than a factor of 2 weaker than after the double pulse.
\newpage
\section*{\textbf{Supplementary Note 6:}\label{sec:Ionic temperature} Partial ionic temperature}

\begin{figure*}[ht!]
\includegraphics[width=0.7\textwidth]{TI_47mHa.ps}
\caption{\label{fig:ion_temp} Ionic temperature as a function of time after a 
laser excitation inducing $S_e = 5.05$ mHa K$^{-1}$. It is decomposed in three phonon 
ranges: acoustic, mixed, optical. As a reference the total ionic temperature 
for the double pulse excitation is shown (light green).}
\end{figure*}
As described in the main text and shown in Suppl. Fig. 3(a), we decomposed the total ionic temperature using 
the PhDOS. For completeness we show here the results for a single-pulse 
excitation to $S_e = 5.05$ mHa K$^{-1}$. Here, the different phonon ranges also do not 
share a common temperature after $1$ ps. However, the difference between the 
ranges is here only of the order of $30$ to $50$ K.

\section*{\textbf{Supplementary Note 7:}\label{sec:quasi stationary} Atomic quasi-stationary state after the second pulse}

\begin{figure*}[ht!]
\includegraphics[width=0.7\textwidth]{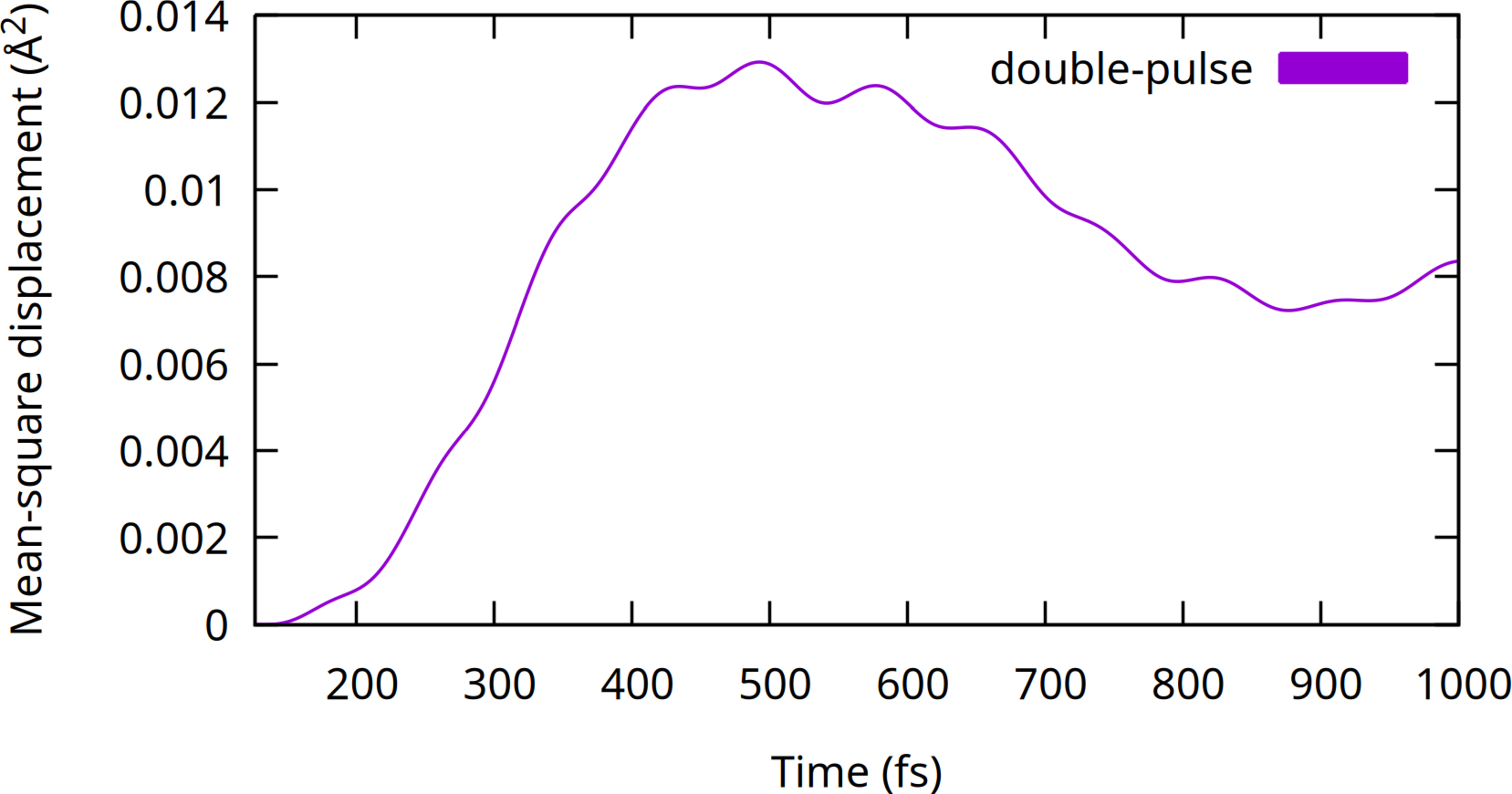}
\caption{\label{fig:quasi_station} Averaged mean-square atomic displacement as function of time computed with respect to the atomic configuration at time $t_2$ of the second pulse. The linewidth, which is plotted equal to the standard deviation, is too small to be seen, reinforcing the laser-induced coherent motion.}
\end{figure*}

The atomic structure remains the same for our simulation time of 1 ps. During this time, 
the atoms move only a small distance from their initial positions. This can be
seen from the time-dependent mean-square atomic displacement, shown in Figure 6 (main text) . 
To quantify this motion, we additionally computed the mean-square atomic 
displacement with respect to the atomic configurations 
$\vec{r}_j(t_2)$ at the time of the second pulse instead of $\vec{r}_{i}^0$ (Supl. Fig. \ref{fig:quasi_station}). Here, we show the results averaged over the ten independent runs used in the main text. As the linewidth, we plotted the standard 
deviation from the mean value. However, no line width is visible, meaning that the
differences between the runs are too small to be seen, which reinforces the coherent nature
of our reported effect. On average, the atoms move only $\sim 0.1$ Å with respect to their
position when the second pulse is applied. As a reference, Figure 6 shows the MSD with respect 
to the ideal diamond-like crystal. There, the atoms move 
on average 0.7 Å ($\sim 0.5$ Å$^2$). This indicates that the atoms are almost perfectly
static at their positions. As a consequence, the whole structure remains throughout our simulation very close to the one at the moment of the second pulse.

\newpage
\section*{\textbf{Supplementary Note 8:}\label{sec:} Estimated temperature range of effect observability}

\begin{figure*}[ht!]
\includegraphics[width=0.7\textwidth]{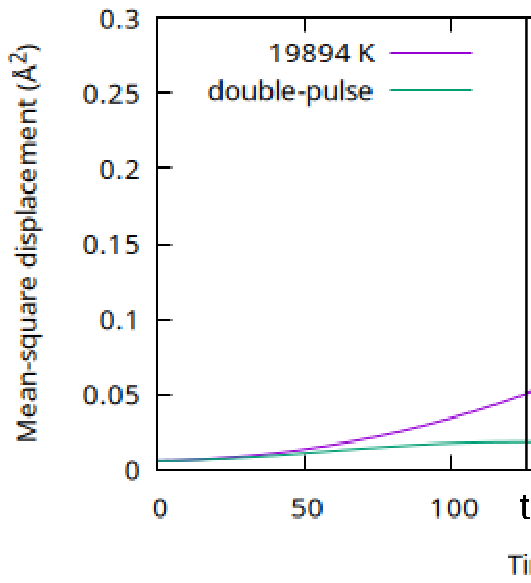}
\caption{\label{fig:MSD_63mHa}Time-dependent mean-square atomic displacement after a single-pulse excitation (purple) and a double pulse excitation (green) inducing an electronic temperature of $19 894$ K. Here, only a single run was performed.}
\end{figure*}

We note that we did not perform a complete study about up to which induced electronic temperature the reported effect of pausing ionic motions is still observable. However, in our optimization process we tried different parameters for the electronic 
temperature. As can be seen in Figure 7, for larger electronic temperatures the effect is not as pronounced as with the chosen parameters but is 
clearly visible. In addition, we performed a single simulation with a laser-induced electronic temperature of $63$ mHa ($19894$ K) (Supplementary Figure \ref{fig:MSD_63mHa}). 
Using the proposed double-pulse scheme at this temperature, we observe that the second 
pulse at $t_2 = 126$ fs does not induce pausing in an optimal manner. However, 
a large difference from the single pulse is observed. Our results therefore indicate that a pausing effect is still significant up to an electronic temperature of 63 mHa (19894 K). As a result, the effect should be observable at least over an electronic temperature range of 2500 K.

\section*{Supplementary References} 
\bibliography{apssamp}